\pdfoutput=1
\documentclass[aps,pra,amsmath,amssymb,superscriptaddress,onecolumn,notitlepage]{revtex4-1}
\usepackage{appendix}
\usepackage{xcolor}
\usepackage{amsmath}
\usepackage{esint}
\usepackage{marvosym}
\usepackage{bm}
\usepackage{amssymb}
\usepackage{layouts}
\usepackage{graphicx}
\usepackage{natbib,hyperref}

\newcommand{\dd}{\mathrm{d}}
\newcommand{\dr}{\mathrm{dr}}
\newcommand{\bdr}{\textbf{dr}}
\newcommand{\q}{\mathrm{q}}
\newcommand{\eff}{\mathrm{eff}}

\newcommand{\be}{\begin{equation}}
\newcommand{\ee}{\end{equation}}

\begin{document}

\title{Generalized Hydrodynamics of the attractive Non-Linear Schroedinger Equation}

\author{Rebekka Koch}
\affiliation{ Institute for Theoretical Physics, University of Amsterdam, PO Box 94485, 1090 GL Amsterdam, The Netherlands}

\author{Jean-S\'ebastien Caux}
\affiliation{ Institute for Theoretical Physics, University of Amsterdam, PO Box 94485, 1090 GL Amsterdam, The Netherlands}

\author{Alvise Bastianello}
\affiliation{ Department of Physics and Institute for Advanced Study, Technical University of Munich, 85748 Garching, Germany}
\affiliation{Munich Center for Quantum Science and Technology (MCQST), Schellingstr. 4, D-80799 M{\"u}nchen, Germany}

\begin{abstract}
We study the generalized hydrodynamics of the one-dimensional classical Non Linear Schroedinger equation in the attractive phase. We thereby show that the thermodynamic limit is entirely captured by solitonic modes and radiation is absent.
Our results are derived by considering the semiclassical limit of the quantum Bose gas, where the Planck constant has a key role as a regulator of the classical soliton gas.
We use our result to study adiabatic interaction changes from the repulsive to the attractive phase, observing soliton production and obtaining exact analytical results which are in excellent agreement with Monte Carlo simulations.
\end{abstract}

\maketitle

\tableofcontents
\section{Introduction}
\label{sec_intro}

The relentless experimental advances in probing strongly correlated quantum matter are hardly overemphasized \cite{Bloch2005,RevModPhys.80.885} and triggered an as much impressive amount of theoretical research.
The one-dimensional world stands out as a fruitful melting pot for experiments \cite{Fukuhara2013,Jepsen2020,PhysRevLett.113.147205,Fukuhara2013_bis,doi:10.1063/1.1704156}, numerical methods and analytical expressions: indeed, particles moving on a line cannot avoid each other, with the consequent enhancement of interactions and correlations.
As a consequence, the fascinating problem of describing nonequilibrium phases of matter in low dimensions is extremely challenging and often far beyond the comfort zone of perturbative methods. Hence, other approaches should be envisaged.

Tensor networks \cite{Schollwoeck2011} are one of the most versatile and powerful numerical methods to access one-dimensional physics, but they come with their own limitations: the nonequilibrium late-time behavior of many-body system is extremely hard to capture, due to its highly entangled nature. If on top of this one also wishes to set aside lattice systems and explore continuous setups, the problem becomes quickly intractable. Other strategies should then be used, for example semiclassical methods or exact solutions: this paper lies at the interface between these two.

A widely known modelling of 1d cold atom experiments is the interacting Bose gas Hamiltonian, which in dimensionless units reads
\be\label{eq_LLH}
\hat{H}=\int \dd x\,  \partial_x \hat{\psi}^\dagger \partial_x \hat{\psi}+c\hat{\psi}^\dagger\hat{\psi}^\dagger \hat{\psi}\hat{\psi}+V(x) \hat{\psi}^\dagger\hat{\psi}\, ,
\ee
with $c$ describing a local pairwise interaction and $V(x)$ being the shallow confining potential and $[\hat{\psi}(x),\hat{\psi}^\dagger(y)]=\delta(x-y)$.
This model hardly needs any introduction due to its widespread use in both experimental \cite{RevModPhys.83.1405,Kinoshita1125,Paredes2004,PhysRevLett.95.190406,Kinoshita2006,PhysRevLett.105.230402,PhysRevLett.107.230404,PhysRevLett.100.090402,PhysRevA.83.031604,PhysRevA.91.043617,PhysRevLett.115.085301} and theoretical  communities.

In experiments, interacting bosons squeezed in elongated traps, either by means of optical lattices or atom chips, are trustfully described by this Hamiltonian. Remarkably, both the trapping potential and the interactions are highly tunable parameters, also in real time. Particularly interesting for us, as we discuss later on, is the possibility of manipulating the interaction by means of Feshbach resonances \cite{PhysRevLett.81.938,PhysRevLett.91.163201,Haller1224,kao2020creating}. 

The ground state and low-energy properties of the repulsive ($c>0$) 1d Bose gas can be efficiently described within bosonization \cite{HALDANE1981153,Cazalilla_2004}, but depending on the question at hand, the usefulness of this approach is at stake: macroscopic changes of the trap or interactions will create excitations high in the spectrum, far beyond bosonization applicability.
In the opposite limit, semiclassical methods take the lead: in the case of large mode occupancy, one can discard the quantum nature of the fields by replacing them with classical variables $\hat{\psi}(x)\to\phi(x)$. The system is now described by the classical Non-Linear Schroedinger (NLS) equation \cite{novikov1984theory,Faddeev:1987ph}
\be\label{eq_NLS}
i\partial_t \phi=-\partial_x^2\phi+2\varkappa|\phi|^2\phi+V(x)\phi(x)\, ,
\ee
where $\phi$ is a classical complex field and $\varkappa$ the classical interaction strength.
This approximation holds when interactions are weak and curiously emerges in two opposite scenarios. On the one hand, it can describe a single macroscopically occupied mode (quasicondensate) and within this setup is often referred as Gross-Pitaevskii equation \cite{pitaevskii2016bose}. Since this description captures a quasicondensate, it can be interpreted as a low temperature limit.

On the other hand, high temperature limits of the Bose gas can be treated with the NLS as well, which now describes a \emph{fluctuating} classical field \cite{doi:10.1080/00018730802564254,PhysRevA.83.043619,PhysRevB.101.245157}. The fact that cold atoms can be described by means of a high temperature fluctuating classical field could appear as a contradiction, but the concept of ``high temperature" must be compared with the energy scales of the 1d experiments, i.e. the interactions and the oscillator frequency of the longitudinal trap, which can be rather weak.
The appeal of semiclassical approximations is self-evident: classical fluctuating fields can be easily simulated with Monte Carlo methods on very large scales and for long times. Besides, external perturbations and noise are easy to implement.

Outside the semiclassical realm, whose applicability is not restricted to the Bose gas, one can probe the true quantum nature of the system with analytical methods. Indeed, the 1d Bose gas (in the absence of the trap) is an outstanding example of integrable model, originally solved by Lieb and Liniger in 1963 \cite{LiebLiniger1963,PhysRev.130.1616} and often referred to as Lieb-Liniger (LL) model after them. When applicable, integrability is powerful: it gives access to exact results directly in the thermodynamic limit, ranging from equilibrium physics \cite{takahashiBook2005} to out of equilibrium setups \cite{Calabrese_2016}. 
As a downside, deriving analytical solutions can be challenging and until recently the effects of soft integrability breaking terms,  such as the longitudinal trap, were hard to include in the theory.

The table turned in 2016 with the proposal of a new hydrodynamic approach to integrability \cite{castroGHD2016,bertiniGHD2016} now dubbed as Generalized Hydrodynamics (GHD), which is built only on the \emph{local} and \emph{instantaneous} integrability of the Hamiltonian, making it the ideal tool to describe several experimental setups. The interested reader can refer to the recent reviews \cite{denardis2021correlation,Bulchandani_2021,bouchoule2021generalized,alba2021generalizedhydrodynamic,bastianello2021hydrodynamics,borsi2021current,cubero2021form}.
Shortly after the original papers, inhomogeneities in the Hamiltonian such as external potentials \cite{SciPostPhys.2.2.014} and time-changes of the interactions \cite{PhysRevLett.123.130602} were readily included. Recently, GHD predictions have been challenged by real experiments showing a spectacular agreement between the two \cite{PhysRevLett.122.090601,doi:10.1126/science.abf0147,PhysRevLett.126.090602,bouchoule2021generalized}.
Noticeably, the very defining properties of integrability is having an extensive number of local conserved quantities that hinder thermalization and cause relaxation to Generalized Gibbs Ensembles \cite{rigol2007relaxation}, as probed in early experiments \cite{Gring1318,Langen207}, and allows for novel nonequilibrium phases. Very interestingly, integrability can stabilize quantum matter in regimes that simply do not exist in its absence: the Bose gas within the attractive phase $c<0$ is an outsdanding example of it.
At equilibrium, the attractive Bose gas is extremely unstable \cite{PhysRevA.73.033611,CASTIN2001419,Tempfli_2008}: in this phase, the gas can form bound states \cite{takahashi2005thermodynamics,PhysRevLett.98.150403,Calabrese_2007,PhysRevA.72.033613} and for any finite temperature the gas collapses in a gigantic molecule consisting of all particles. However, integrability overturns the rules of the game by stabilizing the few-particles bound states and allowing stable nonequilibrium phases.
This has been already realized and discussed in experiments \cite{Haller1224,kao2020creating}, but theoretical results are scarce \cite{PhysRevLett.116.070408,SciPostPhys.1.1.001} and for a good reason: one should initialize the system in a far from equilibrium setting and this is extremely difficult to be analytically addressed.

Fortunately, GHD comes in handy when facing this challenge: in a recent work \cite{KochBastianelloCaux2021}, the three of us proposed to initialize the gas in the attractive regime by means of a slow interaction tuning starting from the repulsive phase. Within each phase, GHD describes the  interaction changes \cite{PhysRevLett.123.130602} and must be supplemented by a description of what happens crossing the non-interacting point $c=0$. When entering the attractive phase, bound states are expected to be produced and we proposed an educated exact ansatz for this process. However, aside the low-density limit, its validity is extremely challenging to be tested due to the absence of independent analytical or numerical computations.

In this work, we address this protocol in the classical Non-Linear Schroedinger equation: the motivations are multifaceted.
First of all, by comparing with large-scale classical simulations we can benchmark the analytical predictions, exploring corrections to adiabaticity and due to integrability breaking corrections, always present in experimental setups.
Secondly, the in se-per-se interest in the thermodynamic description of integrable models is fueled by the advances of GHD studies. 
In contrast with quantum models, where excitations are of fermionic or bosonic nature, classical systems feature either solitonic or radiative modes, obeying Maxwell-Boltzmann or Rayleigh-Jeans statistics.
Probably, the simplest example of classical integrability is the hard rods gas \cite{Doyon_2017}, namely an ensemble of hard-core particles of finite length obeying a (dressed) Maxwell-Boltzmann statistics, thus having solitonic nature. Solitonic modes also appear in the Toda chain \cite{Cao_2019,doi:10.1063/1.5096892} and Landau-Lifschitz spin model \cite{TAKHTAJAN1977235,PhysRevLett.125.070601,10.21468/SciPostPhys.10.4.086}.
In contrast, the thermodynamics of the relativistic sinh-Gordon model \cite{deLucaMussardo2016equilibration,10.21468/SciPostPhys.4.6.045} and the repulsive NLS \cite{10.21468/SciPostPhys.9.1.002} is built upon radiative modes.
Within the attractive phase, the NLS is known \cite{Faddeev:1987ph} to support both radiative and solitonic modes, hence its thermodynamics can be naively expected to be described by a mixture of the two.
Surprisingly, we show that this is not the case: by accessing the classical GHD through a semiclassical limit of the quantum model, we recover only solitonic modes. Benchmarks with ab initio classical simulations confirm the correctness of the procedure.

The paper is organized as follows. Section \ref{Sec_LLmodel} reviews the solution of the quantum Bose gas, its hydrodynamics and slow interaction changes from the repulsive to the attractive phase. Section \ref{sec_classicalNLS} presents the classical NLS, discussing the difficulties of obtaining a hydrodynamic description from purely classical methods and thus building the GHD from semiclassical limits of the quantum model.
Section \ref{sec_numerical} benchmarks the hydrodynamic treatment against ab initio Monte Carlo simulations. Finally in Section \ref{sec_conclusions} we gather our conclusions followed by two short appendixes desribing the numerical methods.

\section{The 1d Quantum Bose gas} \label{Sec_LLmodel}

In this section, we summarize the key points of the exact solution of the Bose gas and its hydrodynamic treatment, which we use as a stepping stone to access the classical NLS.
We start by reviewing the microscopic solution by means of the Coordinate Bethe ansatz and discussing the strings representing bound-states in the attractive phase. We then turn to the thermodynamic limit and conclude with a short summary of GHD
and its application to slow drives from a repulsive to an attractive interaction strength.

\subsection{The many-body wavefunction}
\label{SubSec_BetheAnsatz}

Already in 1963 Lieb and Liniger provided the exact solution of the homogeneous $V(x)=0$ 1d Bose gas \cite{LiebLiniger1963}, which is also referred to as Lieb-Liniger (LL) model.
Integrable models are characterized by the presence of an extensive number of local conservation laws which greatly affects the dynamics. Indeed, the Hilbert space can be understood as a collection of particles undergoing pairwise elastic scatterings.
The rapidities $\lambda_i$ represent the asymptotic momenta of the particles after a scattering event.
The asymptotic states are common eigenvectors $|\{\lambda_i\}^N_{i=1}\rangle$  of the Hamiltonian and all the charges, which act additively on the quasiparticles due to their locality
\be
\hat{\mathcal{Q}}|\{\lambda_i\}_{i=1}^N\rangle=\left(\sum_{i=1}^N q(\lambda_i)\right)|\{\lambda_i\}_{i=1}^N\rangle\, ,
\label{Eq_chargeEV}
\ee
where $q(\lambda)$ is called charge eigenvalue. Notable examples are the number of particles $q(\lambda)\to 1$, the energy $q(\lambda)\to \epsilon(\lambda)=\lambda^2$ and the momentum $q(\lambda)\to p(\lambda)= \lambda$.

The many-body wavefunction is explicitly computed via the Coordinate Bethe Ansatz \cite{LiebLiniger1963} and reads
    \be
    \Psi(\{x_i\}_{i=1}^N;\{\lambda_i\}_{i=1}^N)=\sum_{\mathcal{P}}A(\mathcal{P})\prod_{j=1}^Ne^{i\lambda_{\mathcal{P}_j}x_j}, \quad x_1\leq x_2\leq...\leq x_N\, . \label{Eq_wavefunction}
    \ee

Above, $\mathcal{P}$ are the permutations of rapidities and the coefficients $A(\mathcal{P})$ encode the factorized scattering of the particles. 
In particular, exchanging two rapidities is equivalent to a scattering process,
hence the state picks up a phase. By defining $\Pi_{j,j+1}$ as the permutation that exchanges the rapidities in position $j$ and $j+1$, it holds that $ A(\Pi_{j, j+1}\mathcal{P})=S(\lambda_{\mathcal{P}_j}-\lambda_{\mathcal{P}_{j+1}})A(\mathcal{P})$, with
    \be
S(\lambda)\equiv e^{i\Theta(\lambda)}\equiv\frac{\lambda+ic}{\lambda-ic}\, ,\label{Eq_ScatteringFuncRep}
    \ee
where $S(\lambda)$ and $\Theta(\lambda)$ are the two-body scattering matrix and phase, respectively. Notice, that for any non zero interaction, one has $S(0)=-1$: as a consequence, the wavefunction \eqref{Eq_wavefunction} vanishes if two or more identical rapidities are considered, enforcing a Pauli exclusion principle on the set $\{\lambda_i\}_{i=1}^N$.
As we will see, the scattering phase fully encodes the interacting nature of the problem and it is a key ingredient in its thermodynamic description.
The first step in this direction is to consider a finite system of length $L$: periodic boundary conditions quantize the allowed rapidities through the Bethe equations
	\be
	e^{iL\lambda_j}=\prod_{k\ne j}S(\lambda_j-\lambda_k), \quad j,k=1,...,N. \label{Eq_BetheEQ}
	\ee
Notice that in the non-interacting limit $S(\lambda)=1$ the standard quantization conditions of non-interacting models are recovered $e^{iL\lambda_j}=1$.
In the presence of interactions, the highly non-linear nature of the Bethe equations make their solution extremely challenging.

In the zero density limit ($L\to\infty$ and $N$ fixed) the nature of the solutions can be classified and strongly depends on the sign of the interaction strength.
Within the repulsive phase $c>0$, only real rapidities $\lambda_j$ are admitted, which describe asymptotically free particles.
On the contrary, in the attractive regime $c<0$ bound states are possible and are described by complex rapidities. In fact, complex $\lambda_j$ with the proper imaginary part are associated with a wavefunction that exponentially vanishes for large particle separation \eqref{Eq_wavefunction}.

At the level of the Bethe equations \eqref{Eq_BetheEQ}, a complex rapidity causes $e^{i\lambda_j L}$ to either diverge or vanish when $L\to +\infty$. The zero or divergence must be counterbalanced by a similar behavior in the product of the scattering matrices and this locks the rapiditiy differences to either the zeroes or poles of $S(\lambda)$, giving rise to regular patterns in the complex plane which are called strings.
Rapidities belonging to the same string have the same real part, but are regularly shifted along the imaginary direction $\{\lambda-i\frac{c}{2}(j+1-2a)\}_{a=1}^j$ where $j\in\{1,2,...\}$ is the length of the string. Physically, a string represents a bound state of $j$ particles traveling together with momentum $\lambda$.
The string hypothesis \cite{takahashiBook2005} assumes that in the true thermodynamic limit ($L\to\infty$ with $N/L$ kept fixed), the strings can  still be used to classify the state. Hence, by grouping together the constituents of each string, one can write the Bethe-Takahashi equations 
\be\label{eq_bethe_taka}
e^{i Lp_j(\lambda_j^\alpha)}=\prod_{(k',\alpha')\ne (j,\alpha)}S_{jj'}(\lambda_j^\alpha-\lambda_{j'}^{\alpha'})\, ,
\ee
which constrain the real center of mass $\lambda_j^\alpha$ of the strings where $j$ 
specifies the string species.
The string scattering matrix $S_{jj'}(\lambda)=e^{i\Theta_{jj'}(\lambda)}$ is simply defined by the product of the scattering matrices of all the constituents of the two scattered strings. In terms of the scattering phase, it reads
\begin{align}
    &\Theta_{jk}(\lambda)=(1-\delta_{jk})\theta_{|j-k|}(\lambda)+2\theta_{|j-k|+2}(\lambda)+...+2\theta_{j+k-2}(\lambda)+\theta_{j+k}(\lambda) 
    \label{Eq_ScatteringFuncAtr}\\
	&\text{and }  \theta_{j}(\lambda)=-2\arctan\frac{2\lambda}{cj}\, .
\end{align}
Similarly, one defines the charge eigenvalues of a given string by summing the contributions of its constituents
$q_j(\lambda)= \sum_{a=1}^j q(\lambda-i\frac{c}{2}\left(j+1-2a)\right)$.
In particular, in the case of energy and momentum one has
\be
\epsilon_j(\lambda)=j\lambda^2-\frac{c^2}{12}j(j^2-1)\,,\hspace{4pc} p_j(\lambda)=j\lambda\, . \label{Eq_EnergyMomentum}
\ee

Particularly relevant is the energy eigenvalue from which one can readily extract important information on the stability of the attractive phase. In fact, the attractive ground state is a giant bound state of all its constituents with energy $E_g=-c^2 N(N^2-1)/12$. Remaining in the zero momentum sector, the first excited state is obtained by removing one particle from the molecule and placing it infinitely far apart, hence $E_e=-c^2 (N-1)[(N-1)^2-1]/12$. Thus, the energy difference $E_e-E_g$ diverges in the thermodynamic limit, signaling the impossibility of exciting the system for any finite temperature.

\subsection{The Bose gas in the thermodynamic limit}
\label{SubSec_Thermodynamics}

The thermodynamic limit is attained by sending  the system size to infinity $L\to \infty$ while retaining a fixed density $n=N/L$.
In this perspective, solving the Bethe equations by analytical or numerical methods appears a daunting task which, however, can be avoided in the thermodynamic limit.
In the spirit of the Thermodynamic Bethe Ansatz (TBA) \cite{takahashiBook2005}, the exact details of a microstate defined by a set of discrete rapidities $|\{\lambda_i\}_{i=1}^N\rangle$ are no longer of importance. 
Instead, we pass over  to a coarse-grained description of states in terms of the so-called root densities $\rho_j(\lambda)$.
For the sake of a more compact discussion, we consider the repulsive and attractive case at the same time, hence we introduce a set of root densities $\{\rho_j(\lambda)\}$, one for each string species: the repulsive phase is obtained by retaining only strings with $j=1$.
Concretely, $L\dd\lambda \rho_j(\lambda)$ counts how many solutions to the Bethe-Takahashi equations of the $j^\text{th}$ string specie lay within the interval $[\lambda, \lambda+\dd\lambda]$, thus Eq. \eqref{Eq_chargeEV} in the thermodynamic limit becomes
\be\label{eq_th_ch}
\langle \{\lambda_i\}_{i=1}^N|\hat{\mathcal{Q}}| \{\lambda_i\}_{i=1}^N\rangle \rightarrow L\sum_j\int \dd \lambda\, q_j(\lambda)\rho_j(\lambda) \, .
\ee
The Bethe-Takahashi equations \eqref{eq_bethe_taka} in logarithmic form constrain the allowed set of rapidities 
$2\pi\mathcal{I}_{j,\alpha}=Lp_j(\lambda_j^\alpha)-\sum_{(k',\alpha')\ne (j,\alpha)}\Theta_{j,j'}(\lambda_j^\alpha-\lambda_{j'}^{\alpha'})$, with $\mathcal{I}_{j,\alpha}$ an integer. As a consequence, one introduces the total root density $\rho^t_j(\lambda)\ge \rho_j(\lambda)$ as the maximum allowed occupancy and $\vartheta_j(\lambda)=\rho_j(\lambda)/\rho_j^t(\lambda)\le 1$ is the filling fraction.
Due to the non-linear nature of the Bethe-Takahashi equations, $\rho_j^t(\lambda)$ is explicitly state-dependent and is defined as the Jacobian of the change of coordinates between the rapidities and the set of integers $\mathcal{I}_{j,\alpha}$. In the thermodynamic limit, this is compactly expressed as $2\pi \rho_j^t(\lambda)=(\partial_\lambda p_j)^\text{dr}$, where the dressing operation is introduced for an arbitrary test function  $\tau_j(\lambda)\to \tau_j^\text{dr}(\lambda)$ as 
	\be
    \tau_j^{\text{dr}}(\lambda)=\tau_j(\lambda) -\sum_k\int\frac{\dd\lambda'}{2\pi}\varphi_{jk}(\lambda-\lambda')\vartheta_k(\lambda')\tau_k^{\text{dr}}(\lambda').
    \label{Eq_dressingAtr}
    \ee
Here, the kernel $\varphi_{jk}(\lambda)$ is the derivative of the scattering phase $\varphi_{jk}(\lambda)=\partial_{\lambda}\Theta_{jk}(\lambda)$.
The dressing operation is ubiquitous in TBA calculations, as we will see.
Due to the coarse-grained nature of the root density representation, many states share the same macroscopic description: the Yang-Yang entropy
\be
    \mathcal{S}[\{\rho_j\}_{j=1}^\infty]=\int \frac{\dd\lambda}{2\pi}\sum_j\rho_j^t(\lambda) \left(-\vartheta_j\ln \vartheta_j-(\vartheta_j^t-\vartheta_j)\ln(\vartheta_j^t-\vartheta_j)\right)
    \label{Eq_entropy}
\ee
counts the degeneracy of the macrostate identified by a given root density, in the sense that the number of microscopic states sharing the same root density scales as $\sim  \exp (L\mathcal{S}[\{\rho_j\}_{j=1}^\infty])$. Notice that $\mathcal{S}$ is nothing else than the entropy of ``fermionic particles" with occupancy $\vartheta_j(\lambda)$ and total phase space distribution $\rho_j^t(\lambda)$: the fermionic nature of the Yang-Yang entropy is due to the Pauli principle exclusions enforced on the rapidities.
The entropic contribution is crucial when aiming to describe an ensemble of states, both in equilibrium and non-equilibrium setups.

\bigskip
\bigskip
\textbf{\emph{Equilibrium thermodynamics---}} One of the earliest successes of integrability has been its ability to exactly describe systems at thermal equilibrium, despite the presence of strong interactions. Let us focus on a thermal state $e^{-\beta(\hat{H}-\mu \hat{N})}$ with $\beta$ and $\mu$ the inverse temperature and chemical potential, respectively. The grand canonical partition function $\mathcal{Z}$ is obtained by summing over all  microscopic states. In the thermodynamic limit, one can coarse grain the microscopic summation by replacing it with a path integral over the root densities
\be
\mathcal{Z}=\sum_{N,\{\lambda_i\}_{i=1}^N}e^{-\sum_{i=1}^N(\beta(\epsilon(\lambda_i)-\mu)}\to \int \mathcal{D}\rho_j\,  e^{L\left[\mathcal{S}[\rho_j]-\sum_j\int\dd\lambda \beta(\epsilon_j(\lambda)-\mu)\rho_j(\lambda)\right]}\, .
\ee

In the thermodynamic limit $L\to \infty$, the path integral localizes on the saddle point solution obtained by minimizing the free energy $F=-\mathcal{S}[\{\rho_j\}]+\sum_j\int\dd\lambda\, \beta[\epsilon_j(\lambda)-\mu]\rho_j(\lambda)$. This minimization procedure determines a \emph{unique} choice of  root densities that fully describes the thermal state.
Within the attractive regime, the ground state's instability is also present at finite temperatures, therefore the resulting equations are stable only within the repulsive regime and read \cite{takahashiBook2005}
    \be
    \varepsilon(\lambda)=\beta[\epsilon(\lambda)-\mu]-\int \frac{\dd\lambda'}{2\pi}\varphi(\lambda-\lambda')\log(1+e^{-\varepsilon(\lambda)}) 
    \label{Eq_effectiveEnergyRep}
    \ee
where one conveniently parametrizes the filling fractions through the effective energy $\varepsilon$
	\be
	\vartheta (\lambda)=\frac{1}{1+e^{\varepsilon(\lambda)}}.
	\label{Eq_fillingRep}
    \ee

\bigskip
\bigskip
\textbf{\emph{Generalized Gibbs Ensembles---}} Besides playing a pivotal role in equilibrium, the root densities fully characterize several non-equilibrium protocols. Let us focus on the paradigmatic example of the homogeneous quantum quench \cite{PhysRevLett.96.136801}: here, the system is initialized in a certain state (or ensemble) that is not an eigenstate of the Hamiltonian. Then, the state evolves under unitary evolution and eventually (local) relaxation is attained. The presence of infinitely many local conservation laws greatly constrains the dynamics of integrable models and hinders thermalization.
Indeed, the steady state reached by integrable models after a quantum quench is usually not described by thermal ensembles, but rather by Generalized Gibbs Ensembles (GGE) \cite{rigol2007relaxation} $e^{-\sum_i \beta_i \hat{\mathcal{Q}}_i}$, where in addition to the Hamiltonian, all the relevant charges must be taken into account. Within the thermodynamic limit, GGEs can be described with minor changes of the thermal analysis, by modifying the source term in the TBA integral equations \eqref{Eq_effectiveEnergyRep}: however, there are two major obstructions to this program. 
The first concerns the charges that need to be included in the GGE: in continuum models such as the Bose gas, the expectation values of ultra local charges often feature ultraviolet divergences \cite{davieskorepin,PhysRevB.88.205131}, which must be properly regularized \cite{PhysRevA.91.051602,Bastianello_2017,Essler_2017}. In addition, also in  the absence of ultraviolet divergences as for example in lattice systems, listing all the charges needed in the GGE is challenging and quasi-local conservation laws must be considered \cite{PhysRevLett.115.157201,Ilievski_2016}. 
Fortunately, the explicit construction of the charges is not needed for the thermodynamic description of the GGE. Indeed, GGEs are understood to be in one-to-one correspondence with the root density description \cite{Ilievski_2016_str}: given a set of root densities,  a GGE of the form $e^{-\sum_i \beta_i \hat{\mathcal{Q}}_i}$  exists from which they can be derived.\\
The second challenge to be addressed is determining the final GGE from the initial conditions: a major breakthrough in this direction has been the quench action approach \cite{PhysRevLett.110.257203,Caux_2016} which extracts the root densities from the knowledge of the overlaps between the initial state and the post quench eigenstates. Thanks to this approach, several quench protocols in integrable models have been exactly solved, but its applicability is limited to those initial states where the aforementioned overlap can be computed \cite{PIROLI2017362}.\\
In the case of the Bose gas, the list of these special states hosts a unique element, namely the non-interacting ground state (i.e. the homogeneous Bose-Einstein condensate). The sudden quench from the non-interacting ground state to the repulsive case is among the first quenches in interacting integrable models ever solved \cite{PhysRevA.89.033601} and it has been lately generalized to the case of attractive interactions \cite{PhysRevLett.116.070408,SciPostPhys.1.1.001}, showing bound state formation.
However, despite the appeal of such an exact solution, the instability of the one-dimensional BEC against thermal fluctuations makes the realization of this protocol  in real-life experiments difficult.
In contrast to sudden quantum quenches, slow time-dependent protocols allows for a much greater flexibility in the choice of the initial state: these setups fall within the realm of hydrodynamics, which we now discuss.

\subsection{Generalized Hydrodynamics and bound state formation}
\label{SubSec_GHDquantum}

When a many-body system is subjected to slow and smoothly varying external perturbations, it can be rightfully expected to locally evolve through quasi-stationary states. This is the onset of hydrodynamics: by invoking a separation of time scales, one assumes the system is locally and instantaneously described by a homogeneous stationary state, which then evolves on  macroscopic times through proper continuity equations. For those models that are evolving through a locally integrable Hamiltonian, stationary states are described by GGEs and the hydrodynamics of GGEs is now known as Generalized Hydrodynamics (GHD) \cite{castroGHD2016,bertiniGHD2016}.
Since its introduction in 2016, it has provided tremendous advances in understanding 1d systems out of equilibrium. 
Here, we focus on the hydrodynamic equations at the Euler scale, i.e. we retain only the first order in a gradient expansion: diffusive corrections are known \cite{PhysRevLett.121.160603,DeNardis2019} and play a pivotal role in the late-time thermalization due to the trap-induced integrability breaking corrections \cite{PhysRevLett.125.240604}. The thermalization time scale is proportional to the square of the sloshing period and can be neglected for very smooth traps.
At the Eulerian scale, the hydrodynamics is governed by the following continuity equation
     \begin{equation}
    \partial_t \rho_j+\partial_x(v^\text{eff}_j\rho_j)+\partial_\lambda( F_j^\text{eff}\rho_j)=0\, ,
    \label{Eq_GHDeq}
    \end{equation}
where the space-time dependent root density describes the local GGE and  we suppress the explicit dependence on rapidities, space and time for the sake of a more compact notation. We discuss the attractive and repulsive cases again at once, the latter being obtained by retaining only the first string species $\rho_{j=1}$ in the forthcoming expressions.
The continuity equation describes the phase space density of particles moving with velocity $v_j^\text{eff}$ and experiencing a force $F^\text{eff}_j$: due to the strongly interacting nature of the system. The velocity and force are suitably renormalized by the presence of the surrounding particles.
The form $v^\text{eff}_j=(\partial_{\lambda}\epsilon_j)^{\text{dr}}/(\partial_{\lambda}p_j)^{\text{dr}}$, where the dressing operation is defined in Eq. \eqref{Eq_dressingAtr}, has been proposed in the original papers and then later rigorously proven in Refs. \cite{PhysRevX.10.011054,PhysRevLett.125.070602} (however, see Ref. \cite{bastianello2021fragmentation} for an exception).
The effective force undergoes an analogous dressing operation $F_j^\text{eff}=f_j^{\text{dr}}/(\partial_{\lambda}p_j)^{\text{dr}}$, where $f_j$ is the bare force, which captures the effects of traps \cite{SciPostPhys.2.2.014} and interaction changes \cite{PhysRevLett.123.130602}
\be \label{Eq_bareForcQ}
f_j(\lambda)=-j\partial_x V+\partial_t c\sum_{j'}\int \dd\lambda'\, \partial_c \Theta_{jj'}(\lambda-\lambda')\rho_{j'}(\lambda')\, .
\ee
Above, we assume for simplicity that the interaction changes in time, but not in space.
The hydrodynamic equation allows for evolving the root density within the repulsive or the attractive phase, but does not provide a recipe to connect the two. For example, let us initialize the system in a given state (e.g. thermal) within the repulsive phase $c>0$ and then gently tune the interaction to smaller values until $c=0^+$ is met. Here, the evolution with the repulsive GHD stops and continues from $c=0^-$ within the attractive regime with the initial conditions being determined by the endpoint of the repulsive evolution.
This boundary condition is not contained within the GHD equations presented so far, which must be supplemented by further considerations: in Ref. \cite{KochBastianelloCaux2021}, the three of us proposed an analytical ansatz to accomplish this task (see also \cite{bastianello2021hydrodynamics,Bastianello_2019}).
Our reasoning is based on entropic arguments and on the fact that at $c=0$ bound states have purely real rapidities and are indistinguishable from unbound particles with the same rapidity.
This last statement is expressed by the continuity equation
	\be
	\rho(\lambda)\Big|_{c=0^+}=\sum_j j \rho_j(\lambda)\Big|_{c=0^-}\, ,
	\label{Eq_chargeCons}
	\ee
which, besides being diagonal in  coordinate space (whose dependence we omit) is also diagonal in  rapidity space. 
In the case of an interaction change from attractive to repulsive, the right hand-side is known and uniquely determines the repulsive root density: physically, a bound state of $j$ particles traveling with velocity $v^\text{eff}_j(\lambda)$ melts in $j$ unbound particles traveling with the same velocity $v^\text{eff}(\lambda')=v^\text{eff}_j(\lambda)$. At the non-interacting point, it is easy to check that the last identity implies $\lambda=\lambda'$. On the other hand, when the interaction is tuned from repulsive to attractive, only particles with the same rapidity can form a bound state, since particles with different velocity would drift apart before having the chance to bind.
Notice that in the more interesting case when the attractive phase is approached from the repulsive one, Eq. \eqref{Eq_chargeCons} does not uniquely fix $\{\rho_j(\lambda)\}$ and further considerations must be made. If bound states at $c=0^-$ are indistinguishable from unbound particles, any choice of the bound states population is equally probable: therefore, the thermodynamic limit will select the most probable configuration, obtained by entropy maximization.
Maximizing the Yang-Yang entropy \eqref{Eq_entropy} at the free point under the constraint of particle conservation \eqref{Eq_chargeCons} uniquely determines the attractive root densities through
	\be\label{eq_maxen_ansatz}
	\varepsilon_j(\lambda)= j\omega(\lambda)+\sum_k (2\min(j,k)-\delta_{jk}) \log(1+e^{-\varepsilon_k(\lambda)})\,,
	\ee
where $\omega(\lambda)$ is the rapidity-dependent Lagrange multiplier whose value must be tuned to enforce Eq. \eqref{Eq_chargeCons}. The effective energy $\varepsilon_j$ parametrizes the state through $\vartheta_j(\lambda)=(1+e^{\varepsilon_j(\lambda)})^{-1}$. 

Similar equations already appeared in the literature \cite{PhysRevLett.56.904,PhysRevB.26.2514} in the context of the non-interacting limit of the sine-Gordon field theory and can be analytically solved  \footnote{The analytical solution has been overlooked in Ref. \cite{KochBastianelloCaux2021}}
\be\label{eq_exact_q_men}
\vartheta_j(\lambda)=1-\frac{\sinh(j \omega(\lambda)/2)\sinh((j+2) \omega(\lambda)/2)}{\sinh^2((j+1) \omega(\lambda)/2)}\, .
\ee
As a last step, the Lagrange multiplier $\omega(\lambda)$ can be connected with the repulsive root density through $\rho(\lambda)=\frac{1}{2\pi} \frac{1}{e^{\omega(\lambda)}-1}$, obtaining an explicit solution for the filling fraction.

The GHD equations together with the ansatz \eqref{eq_maxen_ansatz} allow one to access the attractive phase in a controlled manner from a large variety of initial states.
Once $c$ is gently tuned to negative values, the gas enters a highly excited phase featuring bound states of finite size, whose stability is protected by integrability. In contrast to equilibrium thermal states, GGEs obtained by solving \eqref{eq_maxen_ansatz} are stable and the bound state population is exponentially decaying in their size $\vartheta_j(\lambda) \sim e^{-j\omega(\lambda)}$. Of course, the value of $\omega(\lambda)>0$ is fixed imposing Eq. \eqref{Eq_chargeCons}: it decreases monotonously as $\rho(\lambda)|_{c=0^+}$, approaching zero when $\rho(\lambda)|_{c=0^+}$ diverges and leading to instabilities.
Our ansatz is rooted on physical arguments and well-motivated thermodynamic considerations, but ab-initio checks are extremely difficult. Analytical consistency checks are available only in the low density limit, while numerical benchmarks are a daunting task. The state-of-the-art tensor network simulations struggle to capture the dynamics of highly excited and entangled systems, especially in the case where the model is continuous and of bosonic nature (which results in an infinite local Hilbert space). In fact, the appeal of GHD is to provide trustful predictions in a regime beyond any other method.
As we will see in the next section, the same hydrodynamic treatment together with the bound states production ansatz can be transferred to the classical world: there, efficient numerical simulations will confirm the correctness of our ansatz.

\section{The Non-Linear Schroedinger equation}
\label{sec_classicalNLS}

After having reviewed the quantum Bose gas and its hydrodynamic description, we now revert to the classical world and introduce the Non-Linear Schroedinger (NLS) equation Eq. \eqref{eq_NLS}.
The equation of motion can be derived from a classical Hamiltonian
\be
H=\int \dd x\, |\partial_x\phi|^2+\varkappa |\phi|^4+V(x)|\phi|^2\, ,
\ee
by enforcing the Poisson brackets $\{\phi(x),\phi^*(y)\}=i\delta(x-y)$.
From the comparison between the Hamiltonian of the 1d quantum model \eqref{eq_LLH} with that of the classical NLS, the similarities between the two appear self-evident.
Indeed, the homogeneous NLS ($V(x)=0$) is a well-known integrable partial differential equation \cite{novikov1984theory,Faddeev:1987ph}: it features infinitely many local conservation laws affecting the dynamics. In the absence of a Hilbert space, there is not a classical analogue of the Coordinate Bethe ansatz and the NLS is studied with the Inverse Scattering Method (ISM). In contrast, the quantum model can also be studied by means of the Quantum Inverse Scattering Method \cite{korepin1997quantum}, but its discussion is beyond the scope of our work.
The key point is noticing that the NLS equation emerges as a compatibility condition of an auxiliary linear problem
\be\label{eq_auxinv}
\partial_x F(t,x)= U_\lambda(t,x) F(t,x)\,, \hspace{2pc} \partial_t F(t,x)= V_\lambda(t,x) F(t,x)\, ,
\ee
where $F$ is an auxiliary two-dimensional vector and $U_\lambda,V_\lambda$ are $2\times 2$ matrices with an explicit dependence on $\phi$
\begin{subequations}
\begin{align}
U_\lambda &=\sqrt{\varkappa}\left(\phi^*\sigma^++\phi\sigma^-\right)+\frac{\lambda}{2i}\sigma^z \\
V_\lambda & =\left(i\varkappa|\phi|^2+\lambda^2\frac{i}{2}\right)\sigma^z-i \sqrt{\varkappa} \left(\partial_x\phi\sigma^+-\partial_x \phi \sigma^-\right)-\lambda \sqrt{\varkappa}(\phi^*\sigma^++\phi \sigma^-)
\end{align}
\end{subequations}
with $\sigma^{x,y,z}$  the standard Pauli matrices and $\sigma^\pm=(\sigma^x\pm i \sigma^y)/2$.

The consistency of the two differential equations \eqref{eq_auxinv} requires the zero curvature condition $\partial_t U_\lambda-\partial_xV_\lambda+[U_\lambda, V_\lambda]=0$ to hold, the latter is $\lambda-$independent and it is satisfied if and only if the Schroedinger equation \eqref{eq_NLS} holds.
Eventually, we are interested in the emergent thermodynamic properties in setups with finite particle number and energy densities, i.e. where $L^{-1}\int \dd x |\phi|^2$ remains constant in the thermodynamic limit.
However, it is convenient to start our discussion within the zero density limit. An extensive discussion can be found in Refs. \cite{Faddeev:1987ph,novikov1984theory}.

\bigskip
\bigskip
\textbf{\emph{The whole line monodromy matrix---}} In this case, the system is assumed to be infinite with the condition that the wavefunction $\phi(t,x)$ vanishes sufficiently fast (e.g. exponentially) for $|x|\to +\infty$.
Let us define the transfer matrix as the spatial propagator of Eq. \eqref{eq_auxinv}
\be
T(t,x,y)=\text{Pexp}\int_x^y\dd z\, U_\lambda(t,z)\, ,
\ee
where $\text{Pexp}$ is the path-ordered exponential. Thanks to the zero curvature condition, the transfer matrix obeys a simple equation of motion
\be\label{eq_ev_mon}
\partial_tT_\lambda(t,x,y)=V_\lambda(t,x)T_\lambda(t,x,y)-T_\lambda(t,x,y)V_\lambda(t,y)\, .
\ee
Particularly important is the role of the monodromy matrix, i.e. the extension of the transfer matrix to the whole line 
\be
\mathcal{T}_\lambda(t)=\lim_{x\to\infty} e^{\frac{\lambda x}{2i} \sigma^z}T_\lambda(t,-x,x)e^{-\frac{\lambda x}{2i} \sigma^z}\, .
\ee
Above, the additional oscillating terms are introduced in view of the asymptotic behavior $\lim_{|x|\to\infty}U_\lambda(t,x)=\lambda\sigma^z/(2i)$.
Due to the vanishing boundary conditions, one simply has $\lim_{|x|\to +\infty} V_\lambda(t,x)=i\sigma^z\lambda^2/2$. From Eq. \eqref{eq_ev_mon} it immediately follows $\partial_t \text{Tr}[\mathcal{T}_\lambda]=0$ for \emph{any} spectral parameter. As a consequence, an infinite $\lambda-$dependent family of conservation laws is found. Further insight is gained from the matrix elements of the monodromy matrix, which can be parametrized as
\be
\mathcal{T}_\lambda=\begin{pmatrix} a(\lambda) & \text{sign}(\varkappa) b^*(\lambda) \\ b(\lambda) & a^*(\lambda) \end{pmatrix}\, ,
\ee
where $a(\lambda), b(\lambda)$ are entire functions in the $\lambda-$complex plane and play the role of action-angle variables of the integrable system. They in fact obey a very simple equation of motion
\be\label{eq_scat_par}
\partial_t a(\lambda)=0 \, , \hspace{2pc} \partial_t b(\lambda)= -i \lambda^2 b(\lambda)\, .
\ee
Hence, $a(\lambda)$ is itself a conserved quantity.
In principle, the non-linear dynamics of the wavefunction $\phi(t,x)$ is solved, albeit in an implicit way. In practice, one should compute the monodromy matrix from the initial profile, then evolve the scattering parameters according to Eq. \eqref{eq_scat_par} and finally use them to infer back the scattering potential and thus $\phi$. The last step is what is referred to as ``inverse scattering" and passes through the solution of the Gelfand–Levitan–Marchenko linear integral equation. In this respect, the analytical properties of $a(\lambda)$ are crucial.
Indeed, $a(\lambda)$ can be analytically continued in the complex and $a(\lambda)\to 1$ when $\lambda$ is large, leading to the representation
\be\label{eq_arep}
a(\lambda)=\prod_{a=1}\frac{\lambda-\lambda_a}{\lambda+\lambda_a^*}\exp\left[\int_{-\infty}^\infty \frac{\dd \lambda'}{\pi i} \frac{\log|a(\lambda')|}{\lambda'-\lambda-i 0^+}\right].
\ee
Above, $\Im (\lambda_a)>0$ and the complex zeroes are associated with \emph{solitons}, i.e. non-dispersive solutions of the NLS traveling at constant velocity. Solitons are present only for attractive interactions $\varkappa$ and are instead absent in the repulsive case. In contrast with solitons, radiative modes (i.e. dispersive solutions) are present for both the repulsive and attractive case.
Eq. \eqref{eq_arep} has a suggestive interpretation: let us consider the repulsive phase where solitons are absent and expand $\log(\lambda)$ for large values of $\lambda$
\be\label{eq_local_ch_cl}
\log a(\lambda)=i\sum_{n=0}^\infty\lambda^{-n-1}\int_{-\infty}^{\infty}\frac{\dd \lambda'}{\pi}(\lambda')^n\log|a(\lambda')|\, .
\ee
On the one hand, $a(\lambda)$ is an integral of motion for any $\lambda$ in view of Eq. \eqref{eq_scat_par}, hence each term $\mathcal{Q}_n\equiv\int_{-\infty}^{\infty}\frac{\dd \lambda'}{\varkappa}(\lambda')^n\log|a(\lambda')|$ is separately conserved. In addition, it is possible to show that $\mathcal{Q}_n$ are \emph{local} conserved charges \cite{novikov1984theory,deLucaMussardo2016equilibration}. On the other hand, a comparison with the expectation value of the quantum charges in the thermodynamic limit \eqref{eq_th_ch} suggests that $\log|a(\lambda)|$ plays the role of a classical root density \cite{doi:10.1063/1.525509,deLucaMussardo2016equilibration}.
However, a proper analysis requires to consider the system in a finite volume, which we now briefly discuss.

\bigskip
\bigskip
\textbf{\emph{The periodic monodromy matrix---}} When periodic boundary conditions are enforced, the monodromy matrix is defined as the transfer matrix across the finite system $\mathcal{T}^\text{PBC}_\lambda(t)=T_\lambda(t,-L/2,L/2)$. Eq. \eqref{eq_ev_mon} still holds and by virtue of PBC $V_\lambda(t,-L/2)=V_\lambda(t,L/2)$, hence the trace of the monodromy matrix is still a conserved quantity.
There are two main differences to the previous case: first, not all  choices of the spectral parameter are acceptable. In fact, by imposing periodic boundary conditions only values of $\lambda$ such that $\mathcal{T}^\text{PBC}_\lambda(t)=\pm\text{Id}$ are allowed. Secondly, in contrast with the fast decaying case, only the trace of the monodromy matrix is conserved, while the diagonal entries are not. However, this is due to boundary terms that are unimportant in the thermodynamic limit if one focuses only on bulk properties: following the argument of Ref. \cite{doi:10.1063/1.525509,deLucaMussardo2016equilibration}, let us imagine a configuration of the field $\phi$ initially having support on the restricted interval $[-L/2+\ell,L/2-\ell]$. In this case, the monodromy matrix on the whole line coincides with the PBC case and has the same analytic properties of its coefficient. In addition, $a(\lambda)$ is conserved until the evolving field does not reach the boundaries.
This suggests that in the thermodynamic limit one can identify $[\mathcal{T}^\text{PBC}_\lambda]_{1,1}\simeq e^{-i\lambda L/2}a(\lambda)$ with $a(\lambda)$ having the \emph{same} analytic properties of the whole line problem. Focusing on the repulsive case, this observation led to the identification of the classical analogue of the Bethe equations \eqref{Eq_BetheEQ}: indeed, by imposing PBC $\log[\mathcal{T}^\text{PBC}_{\lambda_j}]_{1,1}=\log (\pm 1)=\pi j$ and using the representation Eq. \eqref{eq_arep} (where solitons are absent since $\varkappa>0$), one obtains the set of quantization conditions
\be\label{eq_cl_rep_BA}
\lambda_j-\fint_{-\infty}^\infty \dd \lambda'\frac{2\varkappa}{\lambda_j-\lambda'}\frac{\log|a(\lambda')|}{\varkappa \pi L}=2\pi j L^{-1}\, ,
\ee
with $\fint$ being the principal value regularization of the integral.
These equations have strong analogies with the quantum Bethe equations in the logarithmic form \eqref{Eq_BetheEQ}, where the spectral parameter is identified with the rapidities and the summation over 
rapidities has already been replaced with an integral over the root density. In particular, the classical root density $\rho(\lambda)$ and scattering phase $\Theta(\lambda)$ can be identified as \cite{deLucaMussardo2016equilibration}
\be\label{eq_cl_scatph_rep}
\rho(\lambda)\leftrightarrow\frac{1}{\pi\varkappa L}\log|a(\lambda)|\, , \hspace{2pc} \Theta(\lambda)\leftrightarrow \frac{2\varkappa}{\lambda}\, .
\ee
The proportionality factor between $\rho(\lambda)$ and $\log|a(\lambda)|$ is determined comparing the local charges obtained from Eq. \eqref{eq_local_ch_cl} with the standard normalization of particle density and Hamiltonian. This correspondence leads to a full characterization of the post-quench GGE in the repulsive phase \cite{10.21468/SciPostPhys.9.1.002} in terms of the initial data.
In principle, one could attempt the same strategy in the attractive case as well, but this path appears unpractical or, at least, very challenging. First, solitonic modes complicate the analytical properties of the scattering coefficient and must be properly taken into account. Strictly speaking, a sharp distinction between solitons and radiation exists only in the whole line problem and is blurred out in the periodic case: nevertheless, we still refer to the complex zeroes of $a(\lambda)$ as solitons. Secondarily, in writing Eq. \eqref{eq_cl_rep_BA} one implicitly assumes only real values of the spectral parameter $\lambda_j$. However, in the attractive case one expects complex solutions as well, in analogy with the quantum case and their classification further complicates the approach. Fortunately, one can avoid a solution of the classical problem and indirectly access its thermodynamic description through a proper semiclassical limit of the well-controlled quantum case.

\bigskip
\bigskip
\textbf{\emph{The NLS as the semiclassical limit of the quantum Bose gas---}} Despite the intrinsic quantumness of nature, our daily life experience is based on classical physics: in the large mode occupation limit, quantum systems are trustfully described by classical equations of motion.
Semiclassical methods have a long lasting success story which is not possible to properly credit here, therefore we restrict our attention to the interplay between semiclassical limits and integrability, emphasizing the connection between the NLS and the quantum Bose gas.
At equilibrium, the quantum-classical correspondence has been studied in early days \cite{PhysRevLett.56.904,PhysRevLett.62.708}, but it was brought back into focus through recent interest in non-equilibrium problems. In Ref. \cite{deLucaMussardo2016equilibration} the idea of studying classical GGEs by taking the semiclassical limit of quantum theories has been put forward, which has been later successfully applied to sudden quench protocols in the NLS \cite{10.21468/SciPostPhys.9.1.002}.
Following the advent of GHD, the hydrodynamics of classical integrable models has been successfully derived from their quantum relatives \cite{10.21468/SciPostPhys.4.6.045,PhysRevLett.125.070601}, in particular the hydrodynamics of the repulsive NLS has been proven to be extremely fruitful in benchmarking GHD predictions against ab initio numerical simulations.
Here, we will follow a similar route and tackle the NLS in its attractive phase $\varkappa<0$, which becomes accessible in view of our previous results in the attractive quantum Bose gas \cite{KochBastianelloCaux2021}.
The semiclassical limit is most conveniently accessed by introducing a small parameter $h$ which plays the role of  Planck's constant. The large occupation limit requires the number of particles to diverge $N\propto h^{-1}$, but a finite limit is attained only if the quantum interaction is rescaled to small values $c= h \varkappa$. This is readily seen from a path integral representation of the quantum propagator
\begin{multline}
\langle \psi_j|e^{-iT \hat{H}}|\psi_i\rangle=\int\mathcal{D}\psi\,  e^{i\int_0^T \dd t'\int \dd x\, \frac{i}{2}(\partial_t\psi^\dagger\psi-\psi^\dagger\partial_t\psi)-|\partial_x\psi|^2-c|\psi|^4 }\Bigg|_{\psi(x,T)=\psi_f(x),\psi(x,0)=\psi_i(x)}=\\
\stackrel{\psi=\phi/\sqrt{h}}{=}\int\mathcal{D}\phi\,  e^{h^{-1}i\int_0^T \dd t'\int \dd x\, \frac{i}{2}(\partial_t\phi^\dagger\phi-\phi^\dagger\partial_t\phi)-|\partial_x\phi|^2-\varkappa|\phi|^4 }\Bigg|_{\phi(x,T)=h^{-1/2}\psi_f(x),\phi(x,0)=h^{-1/2}\psi_i(x)}.
\end{multline}
In the $h\to 0$ limit, the path integral localizes on the saddle point, enforcing the classical equations of motion. The whole analysis is then carried out on thermal ensembles and GGEs. In particular, the limit of large occupation modes must be imposed for each conserved charge, leading to the scaling
\be\label{eq_mode_ch}
\langle\mathcal{Q}^\text{quantum}\rangle= h^{-1} \langle\mathcal{Q}^\text{classical}\rangle\, .
\ee
In the next section, we briefly review the semiclassical limit of the quantum TBA and GHD in the repulsive case before moving on with the attractive interactions.

\subsection{The semi-classical limit of the repulsive LL}

In this and in the forthcoming sections, it is convenient to introduce an extra label to tell apart quantum and classical quantities. Therefore, we add the subscript ``\rm{q}" to variables labelling quantum quantities.
For the repulsive phase, we follow Refs. \cite{deLucaMussardo2016equilibration,10.21468/SciPostPhys.9.1.002}. As we anticipated, the classical model features a single root density describing radiative modes.
Let us rewrite the large occupancy scaling of the charges \eqref{eq_mode_ch} using the root density representation 
\be\label{eq_largemode_root}
\int \dd\lambda\, q_{\rm{q}}(\lambda)\rho_{\rm{q}}(\lambda)= h^{-1} \int \dd\lambda \, q(\lambda)\rho(\lambda) .
\ee
The charge eigenvalues of the local charges such  as the density or the energy do not bear any explicit interaction dependence. Hence, one can identify the quantum eigenvalues with the classical ones $q_{\rm{q}}(\lambda)=q(\lambda)$: assuming this identification holds for a complete set of charges, imposing Eq. \eqref{eq_largemode_root} and at the leading order one is led to the scaling $\rho_{\rm{q}}(\lambda)= h^{-1}\rho(\lambda)$, with possible further corrections in $h$.
Indeed, this quantum-classical correspondence turns out to give a finite limit of the whole thermodynamic Bethe ansatz with the already anticipated classical scattering phase \eqref{eq_cl_scatph_rep}.
The classical scattering phase is recovered by the quantum one as
\be\label{eq_scat_lim_rep}
\Theta(\lambda)=\lim_{h\to 0}\frac{1}{h}\left(\Theta_{\rm{q}}(\lambda)+\pi \text{sign}(\lambda)\right)\, .
\ee
In order to keep the standard definition of the total root density, the classical root density is consequentially obtained as $\rho^t(\lambda)=\lim_{h\to 0}(\rho^t_{\rm{q}}(\lambda)-\rho_{\rm{q}}(\lambda))$. With this caveat, the classical dressing is defined in the usual manner employing the classical scattering phase and the identification $2\pi\rho^t(\lambda)=(1)^\text{dr}(\lambda)$ still holds: consistently, the limit of the whole GHD results in the same equations as Eq. \eqref{Eq_GHDeq} where the classical scattering phase replaces the quantum one.
An important difference arises in the integral equations describing thermal states and this is deeply connected to the different statistical properties of the classical NLS when compared with the quantum Bose gas, encoded in Eq. \eqref{eq_scat_lim_rep}. Indeed, the Pauli principle exclusion on the quantum Bethe roots naturally leads to fermionic statistics \eqref{Eq_entropy}, while the radiative modes of the repulsive NLS are expected to follow a Rayleigh-Jeans distribution, which can be thought of as the large occupation limit of bosonic particles. In fact, Eq. \eqref{eq_scat_lim_rep} removes the fermionic component of the scattering phase before sending $h\to 0$: this is strictly connected with the possibility of reformulating the whole quantum TBA in terms of bosonic degrees of freedom \cite{doi:10.1143/JPSJ.54.3727}.
By taking the semi-classical limit of the Yang-Yang entropy \eqref{Eq_entropy}, one finds
\be
\mathcal{S}_{\rm q}\simeq\int\frac{\dd \lambda}{2\pi}\Big\{\rho^t(\lambda) \log \vartheta(\lambda) +(1-\log h ) \rho^t(\lambda)\Big\}\, ,
\ee
that still has a divergent term $\propto \log h$. 
However, by explicitly using the definition of $\rho^t$ it is immediate to see that $\int \dd\lambda \rho^t(\lambda)$ is a state-independent, since 
\be
\int \dd\lambda\, \left\{ \rho^t(\lambda)-\frac{1}{2\pi}\right\}=\int \frac{\dd\lambda}{2\pi}\partial_\lambda \Theta(\lambda-\lambda') \rho(\lambda')=0\, .
\ee
Hence, it can be neglected, leading to the definition of the classical Yang-Yang entropy within the repulsive phase
\be
\mathcal{S}\equiv\int\frac{\dd \lambda}{2\pi}\rho^t(\lambda) \log \vartheta(\lambda)\, .
\ee
Just as in the quantum case, thermal states are obtained by minimizing the free energy, leading to a Rayleigh-Jeans distribution of the mode occupation $\vartheta(\lambda)=1/\varepsilon(\lambda)$ 
\be\label{eq_effen_cl}
\varepsilon (\lambda)=
\beta  \big[\epsilon(\lambda)-\mu\big]- \fint_{-\infty}^{\infty}\frac{{\rm d}\lambda'}{2\pi}\partial_\lambda\Theta(\lambda-\lambda')\log \varepsilon (\lambda')\,.
\ee
Thanks to the equipartition theorem, non-interacting waves on thermal states are distributed according to the inverse of the energy: Eq. \eqref{eq_effen_cl} generalizes this expectation to the presence of interactions. Eq. \eqref{eq_effen_cl} matches the natural expectation inferred by the inverse scattering solution, from which only radiative modes of a single species are expected.
This must be contrasted with the attractive phase we now discuss.

\subsection{The attractive phase}
As we have already discussed, tackling the thermodynamics and hydrodynamics of the attractive phase with the inverse scattering solution appears challenging.
Nevertheless, the semiclassical limit offers a direct path to our goal, whose validity has to be ultimately benchmark against numerical simulations.
The first point to be addressed is the correspondence between the quantum and the classical root densities: we start by noticing that in the semiclassical limit the energy quantization of the strings is expected to merge in a continuum. 
In order to understand the correct scaling, let us focus on the expectation value of a given quantum charge
\be
L^{-1}\langle\hat{\mathcal{Q}}\rangle= \sum_{j=1}^\infty \int \dd\lambda\, q_j(\lambda)\rho_{{\rm q}; j}(\lambda)=\sum_{j=1}^\infty \int \dd\lambda\, \sum_{a=1}^j q(\lambda-ih(j+1-2a)\varkappa/2)\rho_{{\rm q}; j}(\lambda) \sim h^{-1}\, ,
\ee
where we explicated the charge eigenvalue using the string hypothesis as well as the correspondence between the quantum and classical interaction strength  $c=h\varkappa$. By requiring the same scaling $\sim h^{-1}$ for all the charges and assuming their completeness, the same scaling must be independently enforced on each rapidity $\lambda$. Furthermore, the imaginary shifts in the charge eigenvalue are equispaced on a vanishing step $\sim h$. This suggests to replace the summation over the quantum strings with an integral over a continuum variable $s= h j$. Therefore we get
\be\label{eq_41}
\sum_{j=1}^\infty  \sum_{a=1}^j q(\lambda-ih(j+1-2a)\varkappa/2)\rho_{{\rm q}; j}(\lambda)
\to \int_0^\infty \dd s \int_0^{s} \dd s' q(\lambda-i(s-2s')\varkappa/2) h^{-2}\rho_{{\rm q}; s/h}.
\ee
Imposing the scaling $\sim h^{-1}$ of the whole expression suggests the following quantum-classical correspondence
\be\label{eq_cl_q_att}
\rho_{\rm{q}; j}(\lambda)= h \rho_{h j}(\lambda)\, . 
\ee
Following this argument, the attractive phase accommodates for the semiclassical limit in a completely different manner when compared with the repulsive case: instead of macroscopically populating a single root density, it fills a divergent number of particle species (i.e. the bound states), but each of them with a population of order $\mathcal{O}(h)$.
This is a priori not entirely obvious from Eq. \eqref{eq_41}, since the same scaling could have been obtained with a single non-vanishing and divergent summand.
However, Eq. \eqref{eq_41} is not the only requirement to be enforced and Eq. \eqref{eq_cl_q_att} guarantees a finite limit of all the TBA equations, as we now discuss.

Before turning to the total root density, we can first deduce the semiclassical limit of the quantum scattering phase \eqref{Eq_ScatteringFuncAtr}, which explicitly reads
\be\label{Eq_kernelsummand}
    \Theta_{\q;jk}(\lambda)=(\delta_{jk}-1)2\arctan\left(\frac{2\lambda}{\varkappa h|j-k|}\right)-2\arctan\left(\frac{2\lambda}{\varkappa h(j+k)}\right)
    -\sum_{a=1}^{\max(j,k)-1} 4\arctan\left(\frac{2\lambda}{\varkappa h(|j-k|+2a)}\right).
\ee

Notice, that by taking the limit of large strings necessary for attaining the semiclassical limit, the summation diverges as $\sim h^{-1}$, while the first two terms remain $\mathcal{O}(h^0)$. Therefore, we define the derivative of the classical scattering phase as
 $\Theta_{s s'}(\lambda)=\lim_{h\rightarrow 0} h\Theta_{\q; s/ h\,\,s'/ h}(\lambda)$: the first two terms in Eq. \eqref{Eq_kernelsummand} do not contribute in this limit
\be
    \Theta_{ss'}(\lambda)= -2\int_{|s-s'|}^{s+s'}\dd \sigma \arctan\left(\frac{2\lambda}{\varkappa \sigma}\right).
\ee
Using the classical scattering phase, one can write the total root density, the dressing equations and all the GHD equations. Of course, they coincide with the semiclassical limit of the quantum hydrodynamics.
In particular, the quantum total root density scales as $\rho^t_{\q; j}(\lambda)= h \rho^t_{hj}(\lambda)$, with the classical total root density defined as the dressed derivative of the momentum $2\pi\rho^t_s(\lambda)=(\partial_\lambda p_s(\lambda))^\text{dr}$. The classical dressing for an arbitrary test function $\tau_s\to \tau_s^\text{dr}$ is defined through the usual integral equation
\be\label{eq_cl_att_dr}
\tau^\text{dr}_s(\lambda)=\tau_s(\lambda)-\int_0^\infty\dd s'\int \dd \lambda' \varphi_{ss'}(\lambda-\lambda')\vartheta_{s'}(\lambda') \tau_{s'}^\text{dr}(\lambda')\, ,
\ee
with $\vartheta_s(\lambda)=\rho_s(\lambda)/\rho^t_s(\lambda)$ the classical filling function and $\varphi_{ss'}(\lambda)=\partial_\lambda \Theta_{ss'}(\lambda)$. It is worth to mention that $\varphi_{ss'}(\lambda)=\partial_{\lambda}\Theta_{s s'}(\lambda)$ can be explicitly written as
\begin{align}
\varphi_{s s'}(\lambda)= \frac{2}{\varkappa }\ln\frac{(s+s')^2 \varkappa ^2+4\lambda^2}{(s-s')^2 \varkappa ^2+4\lambda^2}. \label{Eq_kernelAtrNLS}
\end{align}
The scattering kernel $\varphi_{ss'}$ \eqref{Eq_kernelAtrNLS} is nothing else than the post-scattering displacement of two colliding classical solitons \cite{Kulish1976} and has recently appeared in connection with superdiffusion in the XXZ spin chain \cite{PhysRevLett.125.070601,Bulchandani_2021}, where the exotic transport has been attributed to excitations of semiclassical nature (see also Ref.\cite{PhysRevB.101.041411,PhysRevB.99.140301,10.21468/SciPostPhys.10.4.086}). In the semiclassical limit, the XXZ spin chain is described by the Landau-Lifschitz model, which is gauge-equivalent to the attractive NLS \cite{Faddeev:1987ph,zakharov1979equivalence}, hence the identical scattering kernels.
The energy and momentum eigenvalues are immediately obtained as $p_s(\lambda)=\lim_{h\to 0} h p_{{\rm q}; s/h}(\lambda)$ and $\epsilon_s(\lambda)=\lim_{h\to 0} h \epsilon_{{\rm q}; s/h}(\lambda)$ and read
\be
p_s(\lambda)=s\lambda\, , \hspace{2pc} \epsilon_s(\lambda)=s\lambda^2-\frac{\varkappa^2}{12} s^3\, .
\ee
Finally, the GHD equations have the usual form $\partial_t \rho_s+\partial_x(v^\text{eff}_s \rho_s)+\partial_\lambda(F^\text{eff}_s\rho_s)=0$, with the following effective quantities  $v^\text{eff}_s=(\partial_\lambda \epsilon_s)^\text{dr}/(\partial_\lambda p_s)^\text{dr}$, $F^\text{eff}_s=f_s^\text{dr}/(\partial_\lambda p_s)^\text{dr}$ and the bare force
\be
f_s(\lambda)=-s\partial_x V+\partial_t\varkappa \int_0^\infty \dd s'\, \partial_{\varkappa}\Theta_{s,s'}(\lambda-\lambda')\rho_{s'}(\lambda')\, .
\ee

We now turn to the semi-classical limit of the Yang-Yang entropy, leading to the definition of the classical entropy within the attractive phase.
On the practical side, this will allow us to access the attractive regime by means of slow interaction changes. Moreover, it will unveil the nature of the excitations described by $\rho_s(\lambda)$.

By using  the quantum-classical correspondence in Eq. \eqref{Eq_entropy}, one finds
	\be
	\mathcal{S}=\int\limits_{\delta(h)>0}^{+\infty} \dd s\int \dd \lambda \, \rho^t_s(\lambda) \left\{\vartheta_s(\lambda)[1-\log  \vartheta_s(\lambda)]-2\log (h)\,\vartheta_s(\lambda)\right\},\label{Eq_divergentEntropy}
	\ee
where a $\sim \log h$ term survives and the role of $\delta(h)$ will be clarified soon.
In contrast with the repulsive phase, where the entropy was pointing out the radiative nature of the excitations, Eq. \eqref{Eq_divergentEntropy} is associated with solitons.
Indeed, consider $N$ indistinguishable classical particles, hence the associated phase space is $1/N!$ and the entropy reads $\mathcal{S}= \log(1/ N!)\sim N(1-\log N)$ , in striking similarity with Eq. \eqref{Eq_divergentEntropy}.
We stress the absence of explicit radiative modes, as one could have naively expected from the inverse scattering method.
Let us now analyze the role of the $\sim \log h$ term: a similar correction appeared in the repulsive case as well, but, as we commented, its contribution was state independent and could be omitted. 
This is not the case in Eq. \eqref{Eq_divergentEntropy} and, in order for the $h\to 0$ limit to give a finite equation of state through entropy maximization, the $\sim \log h$ term should be counterbalanced by another singularity. 
This is the role covered by $\delta(h)$: indeed, as we now discuss, $\vartheta_s$ develops a power law singularity for small $s$ and $\delta(h)$ must be properly tuned to obtain a finite expression. From the semiclassical limit perspective, we are summing over string indexes $j=\{1,2,3,..\}$ hence $s\in\{h,2h,3h,...\}$: therefore, a cutoff $\delta(h)\propto h$ is expected.
We notice that a similar $\log h$ term appeared also in the semiclassical TBA of the XXZ chain \cite{PhysRevLett.125.070601}, further strengthening the connection with the NLS. In that case, the role of the Planck constant was played by an external magnetic field.
In the following, we determine  $\delta(h)$ through a bootstrap approach, asking the maximum entropy approach to guide us towards a finite (and explicitly $h-$independent) formulation of the  bound state production. 

\bigskip
\bigskip
\textbf{\emph{Crossing $\varkappa=0$ and soliton production---}} When crossing over from the repulsive to the attractive phase, the classical problem can be tackled in the same spirit as the quantum Bose gas. 
One can proceed in two different but equivalent ways, either by taking directly the semiclassical limit of the exact quantum solution \eqref{eq_exact_q_men} or through maximization of the classical entropy.
We start by addressing the first case, then we turn back to the classical entropy maximization and discuss the role of the $\log(h)$ term appearing in Eq. \eqref{Eq_divergentEntropy}.
The semiclassical limit of Eq. \eqref{eq_exact_q_men} is extremely simple to be taken: by enforcing the $h-$scaling discussed in the previous section, one reaches the simple expression
\be\label{eq_max_en_cl_fill}
\vartheta_{s}(\lambda)= \frac{1}{(2\pi \rho(\lambda))^2} \frac{1}{4\sinh^2( \frac{s}{4\pi  \rho(\lambda)})}\, .
\ee
Notice that the filling fraction diverges for small $s$ as $\vartheta_s(\lambda)\sim 1/s^2$, regardless of the value of $\rho(\lambda)$.
This divergence is the reason behind the appearance of $\log(h)$ in the classical entropy, as we now discuss.
By enforcing that the two phases should be indistinguishable when approaching the non-interacting point, one reaches the semiclassical analogue of \eqref{Eq_chargeCons}
\be\label{Eq_chargeCons_cl}
\rho(\lambda)\big|_{\varkappa=0^+}=\int_0^\infty \dd s\, s\rho_s(\lambda)\big|_{\varkappa=0^-}\, .
\ee

This continuity equation does not suffice to determine the population of each string, which is in turn determined by the constrained maximization of the classical entropy. 
In the limit $\varkappa\to 0^-$, the kernel of the dressing equation becomes diagonal in  rapidity space 
	\be
	\lim_{\varkappa\rightarrow 0^-}\varphi_{s,s'}(\lambda)=	\lim_{\varkappa\rightarrow 0^-}\frac{2}{ \varkappa}\ln\frac{\varkappa^2\left(s+s'\right)^2+4\lambda^2}{\varkappa^2\left(s-s'\right)^2+4\lambda^2}=-4\pi \min(s,s')\delta(\lambda)\, ,
	\ee
As a consequence, the definition of the total root density is also diagonal in  rapidity space and the entropy can be maximized for each rapidity independently leading to the equation
\be\label{eq_vareps_cl}
\varepsilon_s(\lambda)=s\omega (\lambda) +2\log h +\int_{\delta(h)}^\infty \dd s' 2\min(s,s')e^{-\varepsilon_{s'}(\lambda)}\, .
\ee
Where we used the natural parametrization $\vartheta_s(\lambda)=e^{-\varepsilon_s(\lambda)}$.
The $\lambda-$dependent Lagrange multiplier $\omega(\lambda)$ is of course determined by enforcing Eq. \eqref{Eq_chargeCons_cl}.
These integral equation can develop singular solutions for $s\to 0$: this is precisely what happens in Eq. \eqref{eq_max_en_cl_fill} and the divergence must be singled out before taking the $h\to 0$ limit.
In view of Eq. \eqref{eq_max_en_cl_fill}, it is useful to introduce the reparametrization
\be\label{eq_reg_eps}
\varepsilon_s=2\log s+s\bar{\varepsilon}_s\, ,
\ee
where $\bar{\varepsilon}_s$ remains finite as $s\to 0$. Then, the domain $[\delta,+\infty]$ in the integral of Eq. \eqref{eq_vareps_cl} is split into two parts $[\delta,C]\cup[C,+\infty]$. The constant $C$ will be later sent to zero and it is meant to be small enough such that $s\bar{\varepsilon}_s\simeq 0$ for $s\in[\delta,C]$.
In Eq. \eqref{eq_vareps_cl} we now focus on $s\in[\delta,C]$ and use Eq. \eqref{eq_reg_eps}

\be\label{eq_singular_epsilon}
2\log s+s\bar{\varepsilon}_s=s\omega (\lambda) +2\log h +\int_{\delta(h)}^C \dd s' 2\frac{\min(s,s')}{s'^2}+
\int_{\delta(h)}^C \dd s' 2\min(s,s')\frac{e^{-s'\bar{\varepsilon}_{s'}(\lambda)}-1}{s'^2}+s\int_{C}^\infty \dd s' 2\frac{e^{-s'\bar{\varepsilon}_{s'}(\lambda)}}{s'^2}\, .
\ee
In the above, the last integral is obviously finite. The first integral on the second row is not only finite, but it vanishes as $C\to\delta\to 0$. Hence, the singularity is entirely due to the integral in the first row $
\int_{\delta(h)}^C \dd s' 2\frac{\min(s,s')}{s'^2}=2\log(s/\delta)+2s\left(\frac{1}{s}-\frac{1}{C}\right)= 2\log(s/\delta(h))+2+\mathcal{O}(s)$. By matching the singular and $\mathcal{O}(s^0)$ terms in Eq. \eqref{eq_singular_epsilon}, we are lead to $\log\delta(h)=\log h +1$. The next natural question concerns the effects of the singular filling $\vartheta_s$ on the behavior of the total root density and, in general, of dressed quantities.
A similar analysis of the $s\sim 0$ behavior can be performed, showing that the total root density is not only finite, but it also quadratically vanishes (see next subsection) as $s\to 0$: this is crucial, since it guarantees the root density $\rho_s(\lambda)=\rho_s^t \vartheta_s(\lambda)$ to remain finite, thus giving finite expectation values of the charges. 
Using the identity $\log\delta=\log h+1$, one can obtain a finite integral equation for $\bar{\varepsilon}$ and explicitly take the $h\to 0$ limit. Indeed, by using $\int_{\delta(h)}^\infty \dd s' 2\frac{\min(s,s')}{s'^2}=2\log(s/\delta(h))+2$, we can rewrite Eq. \eqref{eq_vareps_cl} as
\be\label{eq_max_en_nns}
\bar{\varepsilon}_s(\lambda)=\omega(\lambda)+\int_{\delta(h)}^\infty \dd s' \frac{2\min(s,s')}{s}\frac{e^{-s' \bar{\varepsilon}_{s'}(\lambda)}-1}{s'^2}\, .
\ee
The integrand is now finite in zero and the limit $\delta(h)\to 0$ can be safely taken.
As in the quantum case, this integral equation admits an exact analytical solution which, with the identification $\rho(\lambda)=\frac{1}{2\pi \omega(\lambda)}$, is equivalent to the result \eqref{eq_max_en_cl_fill} derived as the semiclassical limit of the quantum filling.
\begin{figure}[t]
\begin{center}
\includegraphics[width=0.4\textwidth]{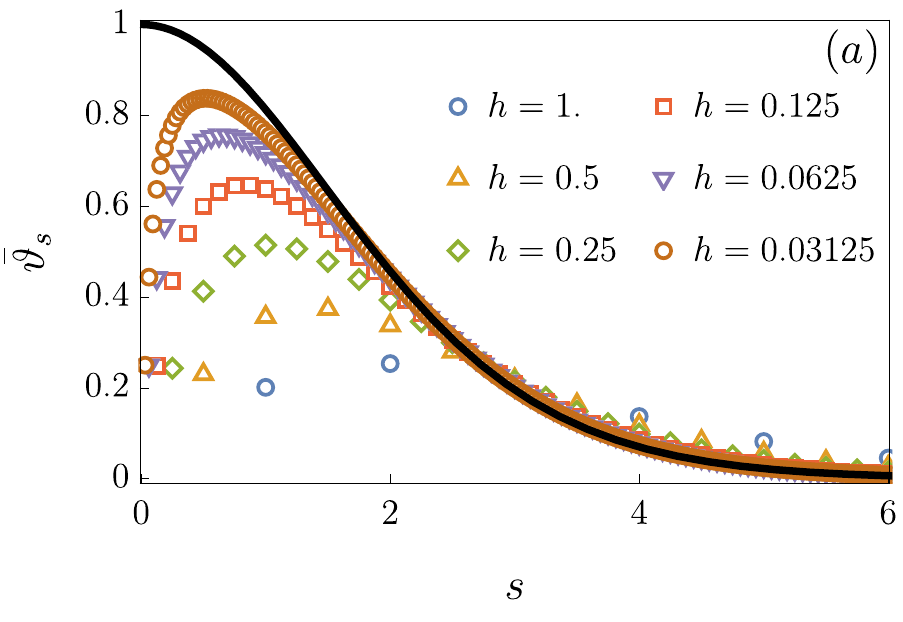}\hspace{2pc} \includegraphics[width=0.4\textwidth]{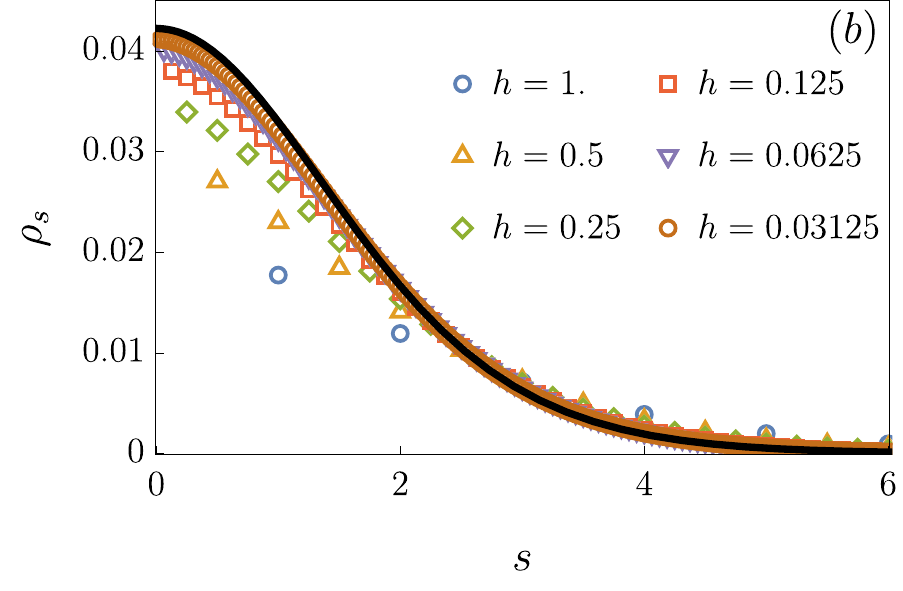}
\caption{\label{fig_maxentr}
We compare the solution of the classical filling and root densities against the quantum case for finite $h$. We proceed as it follows. Since the bound state production is diagonal in  rapidity space, we fix a certain rapidity and give  the value of $\rho(\lambda)\big|_{\varkappa=0^+}$ as an input, setting it to $0.1$ in this figure. Then, the quantum root density is initialized as $\rho_{\q}(\lambda)\big|_{c=0^+}=h^{-1}\rho(\lambda)\big|_{\varkappa=0^+}$ and filling functions are shown for different values of $h$.
Panel $(a)$: the non singular part of the classical filling $\bar{\vartheta}_s$  \eqref{eq_barvarth} (solid line) is compared against the quantum data $( h j,  h^2 \vartheta_{\q; j})$ (symbols).
Panel $(b)$: the classical root density $\rho_s$ (solid line) is compared against the quantum data $(h j, h^{-1} \rho_{\q; j})$ (symbols, obtained by numerically solving the quantum and classical dressing equations).}
\end{center}
\end{figure}

In Fig. \ref{fig_maxentr} we compare the semiclassical scaling with the quantum result for small values of $h$ and study the convergence. In the case of the filling fractions, we compare the analytical solutions while the correspondent root densities are obtained by numerically solving the dressing equations.
Overall, we experience that a large number of quantum strings are needed to reproduce the classical result, especially in the case of the filling function. In contrast, the classical dressing equations can be efficiently discretized with a far lower number of points (in this example, excellent convergence is already obtained with $\sim 20$ points). This stresses the need of solving directly the classical GHD rather than numerically considering the semiclassical limit of the quantum case.

\bigskip
\bigskip
\textbf{\emph{A non-singular reparametrization---}} The analysis of the singularities we provided suggests to redefine the total root density and filling by singling out the small $s$ behavior. The same procedure is also extremely useful for computing quantities appearing in the GHD equations. 
Hence, we introduce the following definitions that are also valid  for finite interactions $\varkappa<0$.
\be\label{eq_barvarth}
\bar{\vartheta}_s(\lambda)\equiv s^2 \vartheta_s(\lambda)\, ,\hspace{2pc}\bar{\rho}_s^t(\lambda)\equiv s^{-2}\rho_s^t(\lambda)\, .
\ee
Furthermore, we define the regularized dressing on a test function $t_s(\lambda)\to t_s^{\bdr}(\lambda)$ as the solution of the following integral equation
\be\label{eq_drbar}
s t_s^{\bdr}(\lambda)=t_s(\lambda)-\int_0^\infty \dd s' \int\dd\lambda' s^{-1}\varphi_{ss'}(\lambda-\lambda')\bar{\vartheta}_{s'}(\lambda') t_{s'}^{\bdr}(\lambda')\, .
\ee
Notice that the limit $\lim_{s\to 0} s^{-1}\varphi_{ss'}(\lambda)$ is finite and non-vanishing for any interactions. As a consequence, if the source term does not have singularities $t_s(\lambda)\sim \mathcal{O}(s^0)$ as $s\to 0$, then $t_s^{\bdr}(\lambda)$ is expected to be finite as well, which is easily numerically checked.
The definition Eq. \eqref{eq_drbar} leads to the following correspondence between the regularized and standard dressing \eqref{eq_cl_att_dr}
\be\label{eq_fills_nos}
\tau_s^\text{dr}(\lambda)= s^2 (s^{-1}\tau_s)^{\bdr}
\ee
In particular, this implies $\bar{\rho}_s^t(\lambda)=(1/(2\pi))^{\bdr}_s(\lambda)$, hence $\bar{\rho}^t_s(\lambda)$ remains finite as $s\to 0$, while $\rho^t_s(\lambda)$ quadratically vanishes, as anticipated.
Due to the $\sim s$ behavior of the energy and bare force, $(s^{-1}\partial_\lambda\epsilon_s)^{\bdr}$ and $(s^{-1}f_s)^{\bdr}$ are finite as well. Hence, the effective velocity and force can now be conveniently computed in terms of a ratio of non-singular functions.

\section{Hydrodynamics and microscopic simulations}
\label{sec_numerical}

One of the major advantages that classical field theories have in contrast to quantum ones is the possibility of numerically accessing arbitrarily long times, hence we are in the perfect position to test our hydrodynamic predictions and quantify the corrections.
In particular, classical simulations can be performed through Monte Carlo methods: the initial field configurations are randomly sampled from a known probability distribution. Then, the initial conditions are deterministically evolved through the equations of motion and the desired observables are computed. Finally, the average over the initial conditions is taken.
Thermal states within the repulsive phase are probably the most natural choice for the initial conditions and can be efficiently simulated by means of Metropolis-Hasting methods \cite{doi:10.1063/1.1699114,10.1093/biomet/57.1.97,doi:10.1080/00031305.1995.10476177}. However, as we will discuss in the following, other choices of initial GGEs are also plausible and easy to simulate.
We perform the standard microscopic numerical simulations following Refs. \cite{10.21468/SciPostPhys.9.1.002,PhysRevLett.125.240604}. For the interested reader, a short summary is reported in  \ref{app_micro_num}.
When numerically solving the GHD equations, it turns out that the root density is not the most convenient variable. Already in the seminal papers \cite{castroGHD2016,bertiniGHD2016} the GHD equations have been rewritten in terms of the filling fraction where convective form is recovered. Within the attractive phase, the GHD equations in terms of the filling fraction are
\be\label{eq_ghd_fill}
\partial_t \bar{\vartheta}_s+v^\text{eff}_s\partial_x \bar{\vartheta}_s+F^\text{eff}_s\partial_\lambda \bar{\vartheta}_s=0\, ,
\ee
and an analogous expression is reached within the repulsive phase, where of course the string index is absent. Notice that, importantly, it is convenient to use the rescaled filling fraction defined in Eq. \eqref{eq_barvarth} to ensure the regularity of the expression for $s\to 0$. Similarly, in the attractive phase the effective velocity and force are computed using the modified definition of the dressing introduced in Eqs. (\ref{eq_drbar}-\ref{eq_fills_nos}). Showing the equivalence between the GHD equations \eqref{eq_ghd_fill} and the canonical version in terms of the root density requires to manipulate the integral equations defining the dressing operation. We choose to not report here these lengthy, but standard calculations: the interested reader can find them in the seminal papers \cite{castroGHD2016,bertiniGHD2016} (see also Ref. \cite{PhysRevLett.123.130602} for the presence of forces) and straightforwardly apply them to our classical case as well. 
Once the GHD equations are written in the form Eq. \eqref{eq_ghd_fill}, they can be seen as infinitesimal translations in the $(x,\lambda)$ plane and solved by the method of characteristics \cite{PhysRevLett.123.130602,mller2020introducing}: we provide a short summary in \ref{app_ghd_num}.
In principle, the GHD equations \eqref{eq_ghd_fill} can be used to study interaction changes in fully inhomogeneous settings, that are for example given by trapping potentials. In practice, we experience that the need of discretizing the continuum set of strings of the attractive phase places a major hurdle on the numerical simulations. For this reason, and since we are mostly interested in observing the bound state production due to time-dependent interaction changes,  we focus here on the homogeneous case. Nevertheless, a more careful optimization of the GHD code would make  the fully inhomogeneous case accessible as well.
Notice that, within the homogeneous case, the $\partial_t \varkappa$ term appearing in the effective force can be singled out and absorbed with a change of variables. Thernefore, the GHD evolution can be parametrized in terms of the interaction only. In the following, we use this convention.
The comparison between hydrodynamics and microscopic simulations passes through the expectation value of observables. As we discussed, in integrable models the expectation values of conserved quantities are the most easy ones to be accessed, leading to the density of particles and energy as the natural choices. However, the particle density is trivially conserved throughout the protocol and is thus not informative at all. On the other hand, the expectation value of the energy density diverges on classical thermal states: this is due to the well known ultraviolet catastrophe of black body radiation, which led to the discovery of quantum mechanics.
Indeed, at very high momenta one can discard the role of interactions in Eq. \eqref{eq_effen_cl} obtaining $\rho(k)\sim (\beta \epsilon(k))^{-1}$ and causing the energy density to diverge $\int \dd k\, \epsilon(k) \rho(k)\to \infty$. For this reason, we focus on the expectation value of the correlated density $\langle |\phi(x)|^4\rangle$, which is finite and undergoes a non-trivial evolution. In the quantum case, its expectation value can be computed on arbitrary GGEs by means of a generalization of the Hellmann-Feynmann theorem (see eg. \cite{KochBastianelloCaux2021}). From this point onwards, the semiclassical limit is readily taken leading to the classical expressions. In particular, one has
\begin{eqnarray}
	\langle |\phi|^4\rangle &=\int\frac{\dd\lambda}{2\pi} 2\lambda\vartheta (\lambda)f ^{\dr}(\lambda) , \hspace{2pc} &\varkappa>0\\
	\langle |\phi|^4\rangle
&=\int \dd s\dd \lambda\Big\{-\rho_s(\lambda)\frac{\varkappa }{6}s^3+\frac{2s\lambda}{2\pi}\vartheta_s (\lambda) f_s^{\dr}(\lambda)\Big\}  \hspace{2pc} &\varkappa<0.
\end{eqnarray}
\begin{figure}[t!]
\begin{center}
\includegraphics[width=0.8\textwidth]{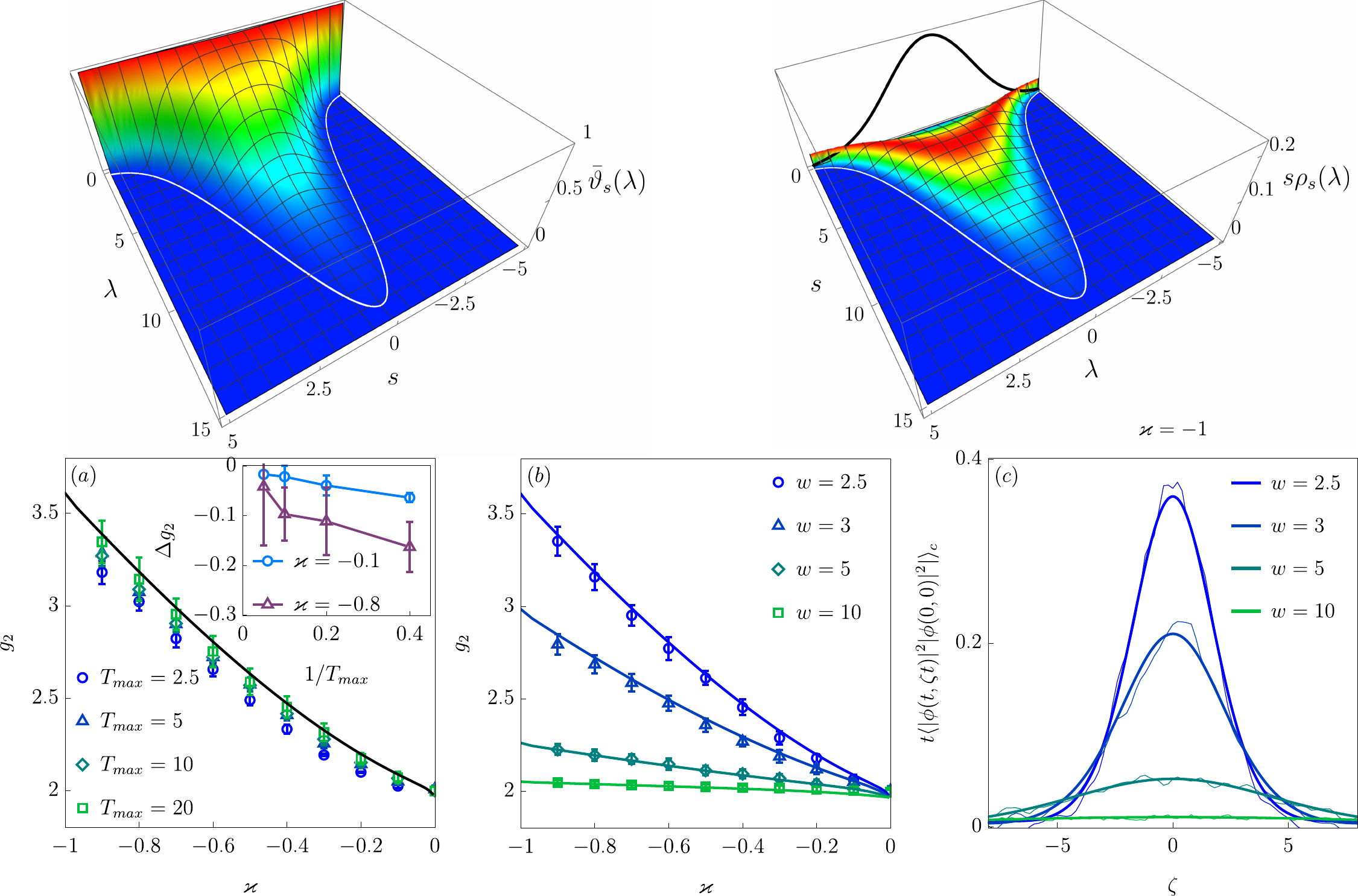}
\caption{\label{fig_gauss} We initialize GGEs in the form of Eq. \eqref{eq_GGE_gauss} at the noninteracting point $\varkappa=0^-$ and drive it into the attractive regime. Top: filling fraction (left) and $s\rho_s$ at $\varkappa=0^-$ for $w=2.5$ in Eq. \eqref{eq_GGE_gauss}. For comparison, $\rho(\lambda)|_{\varkappa=0^+}$ (black curve) is also provided.
Panel $(a)$: corrections to adiabaticity to $g_2=\langle|\phi|^4\rangle/(\langle|\phi|^2\rangle)^2$ are considered for $w=2.5$ (inset, scaling of the microscopic data to GHD as a function of $T_{max}$).
Panel $(b)$: $g_2$ for different initial states (microscopic simulation with $T_{max}=20$).
In $(a)$ and $(b)$: solid line: GHD; markers: Monte Carlo.
Panel $(c)$: density-density Eulerian correlation functions at the endpoint of the protocol (i.e. $\varkappa=-1$) for different initial states (thick curves: GHD; thin curves: Monte Carlo). The correlator is measured at the end of the interaction ramp ($T_{max}=20$), by holding constant the interaction for a time $t$ before measuring. Here, we already attained a good Eulerean scaling by choosing $t=2$.
See the main text for further comments,  \ref{app_micro_num} and \ref{app_ghd_num} for  Monte Carlo and GHD simulations respectively.}
\end{center}
\end{figure}
In the second line we can use the non-singular parametrization we introduced by noticing $\vartheta_s(\lambda) f^\dr_s(\lambda)=\bar{\vartheta}_s(\lambda)(s^{-1}f_s)^\bdr$.
The amount of information of the evolving state captured by the correlated density is of course limited and a better characterization of the state though further measurement is desirable.
Within the repulsive case, close analytical expression for the exact probability distribution of the density $|\phi|^2$ have been recently derived \cite{10.21468/SciPostPhys.9.1.002}. Consequentially, all the momenta $\langle|\phi|^{2n}\rangle$ can be computed on arbitrary GGEs (see also Refs. \cite{PhysRevLett.120.190601,Bastianello_2018} for the quantum case). Moreover, the root density itself could be determined from the microscopic simulations thanks to the correspondence \eqref{eq_cl_scatph_rep}.
However, to the best of our knowledge, analogous results are not available within the attractive realm, hence we turn our attention to different observables in the form of two-point correlation functions.
Indeed, GHD allows for an exact computation of two-point correlation functions $\langle \mathcal{O}(t,x)\mathcal{O}(t',y)\rangle$ on the Eulerian scale (i.e. leading order in the large space and time separation limit), provided the expectation value of $\langle \mathcal{O}\rangle$ on arbitrary GGEs is known.
This result has first been obtained for the time independent and homogeneous GGEs \cite{10.21468/SciPostPhys.5.5.054} and was tested for the first time on the classical Sinh Gordon model \cite{10.21468/SciPostPhys.4.6.045}. Later, expressions for inhomogeneous and time dependent GGEs became available \cite{10.21468/SciPostPhysCore.3.2.016} (see also \cite{denardis2021correlation}). For simplicity, here we focus on the two-point correlation of the density on constant GGEs
\be\label{eq_corr_euler}
t\langle\mathcal{O}(t,x)\mathcal{O}(0,0)\rangle_\text{c}=\int\dd s\,\dd\lambda\,\, \delta \big(x/t-v_s^\text{eff}(\lambda) \big)\rho_s(\lambda)\sigma_s(\lambda) [V_s^\mathcal{O}(\lambda)]^2\, ,
\ee
where $\langle...\rangle_\text{c}$ stands for the connected part of the correlator and the integration over the string variable is obviously omitted in the repulsive case $\varkappa>0$.
Eq. \eqref{eq_corr_euler} holds for large space-time separations and after averaging over mesoscopic fluid cells. Notice the scaling form of the right hand term as a function of the ray $\zeta\equiv x/t$.
The function $V^\mathcal{O}$ carries the information of the observable of interest and it is defined as $\rho^t_s(\lambda) V^\mathcal{O}_s(\lambda)=\frac{\delta \langle \mathcal{O}\rangle}{\delta \vartheta_s(\lambda)}$. Notably, in the case where $\mathcal{O}$ is the local density of a conserved charge, the simple result $V^\mathcal{O}_s(\lambda)= q_s^\dr(\lambda)$ holds true. Lastly, $\sigma_s(\lambda)$ is a statistical factor determined by the nature of the excitations and it is different if quantum particles (bosons or fermions) or classical radiation or solitons are considered \cite{10.21468/SciPostPhys.5.5.054}. In our case, one has
\be
\sigma(\lambda)=\vartheta(\lambda)\, ,\hspace{1pc} \varkappa>0\,;\hspace{4pc}
\sigma_s(\lambda)=1\, ,\hspace{1pc} \varkappa<0\, .
\ee
Of course, the Eulerian correlation functions can also be accessed by first writing the quantum expression and then taking the semiclassical limit, consistently obtaining the same result.
We now benchmark our analytical predictions against microscopic simulations. For numerical reasons clarified later on, rather than starting from a repulsive thermal state and drive the interactions in the attractive phase, we start by engineering suitable GGEs at the non-interacting point $\varkappa=0$ of the form
\be\label{eq_GGE_gauss}
\rho(\lambda)\big|_{\varkappa=0^+}=\frac{1}{w\sqrt{\pi}} e^{-\lambda^2/w^2}\, .
\ee
\begin{figure}[t!]
\begin{center}
\includegraphics[width=0.8\textwidth]{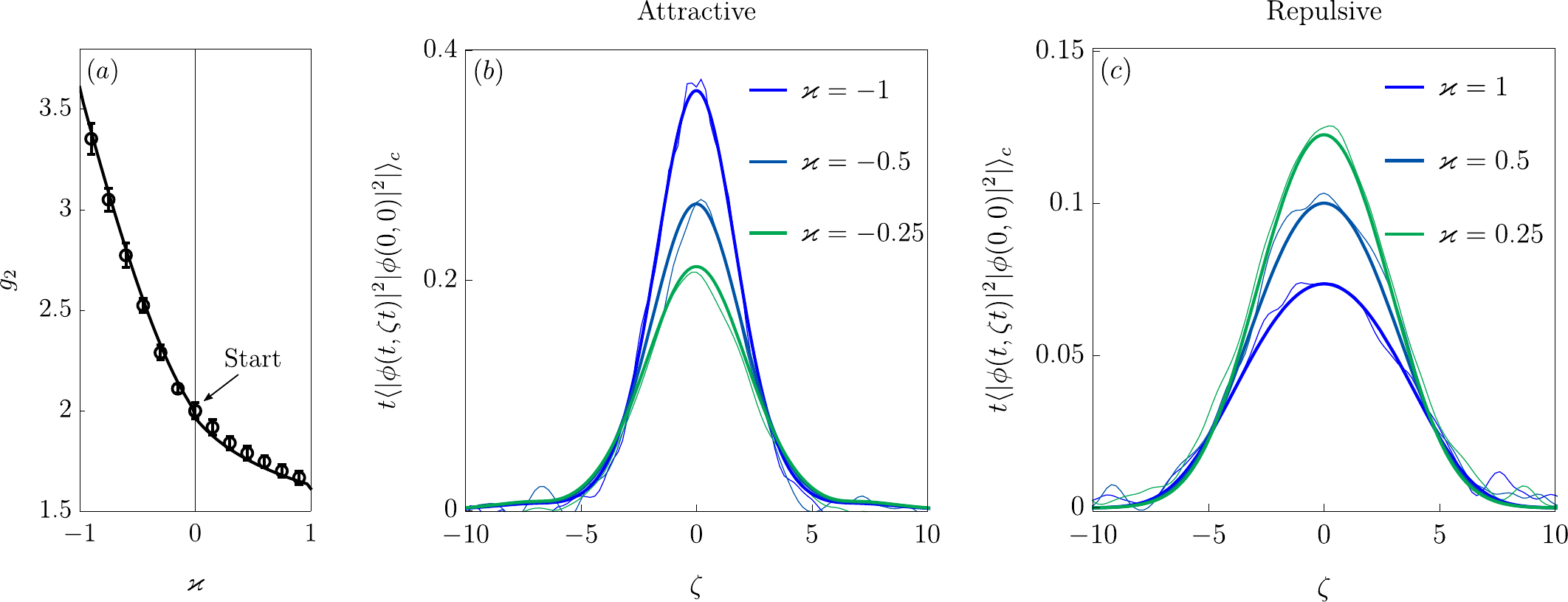}
\caption{\label{fig_gauss_rep} We initialize the system in a non-interacting GGE of the form Eq. \eqref{eq_GGE_gauss} with $w=2.5$ and adiabatically drive it within the repulsive and attractive phase ($T_{max}=20$ in microscopic simulations). Panel $(a)$: $g_2$ is measured (solid line: GHD; markers: Monte Carlo).
Panel $(b)$ and $(c)$: Eulerian correlation function in the attractive and repulsive phase respectively (thick curves: GHD; thin curves: Monte Carlo; holding time $t=2$).
See main text for further comments.
}
\end{center}
\end{figure}
With this convention, the total density of particles is fixed to unity and the parameter $w$ is tuned to address different initial states. Notice that the root density is  decaying fast for $\lambda\to \infty$: this is not the case for thermal ensembles, which makes the discretization of the GHD equations more difficult. GGEs in the form Eq. \eqref{eq_GGE_gauss} are easy to be implemented in the microscopic simulations, see \ref{app_micro_num}.
In Fig. \ref{fig_gauss} we investigate several protocols obtained by gently driving the system within the attractive phase $\varkappa<0$ starting from states \eqref{eq_GGE_gauss}.
We consider linear ramps of the interaction $\varkappa(t)=-t/T_{max}$ and start by investigating the convergence to GHD as $T_{max}$ is increased (Fig. \ref{fig_gauss} $(a)$). We focus on the correlated density $g_2=\langle |\phi|^4\rangle/(\langle|\phi|^2\rangle)^2$ and observe a linear convergence in the ramp time $1/T_{max}$, within the numerical error (see inset).
Corrections to adiabaticity are of different nature: higher derivative terms in the GHD equations \eqref{eq_ghd_fill} are expected and a second order contribution $\sim \partial_t^2$ would lead to $\sim T^{-1}_{max}$ corrections. Besides, finite-time protocols affect the bound state recombination happening at $\varkappa=0^-$. Also in this case, one can envisage $\sim T^{-1}_{max}$ corrections in view of the forthcoming argument. Following the quantum case discussed in Ref. \cite{KochBastianelloCaux2021}, immediately after crossing $\varkappa=0^-$, bound state recombination is allowed as long as the typical size of the bound state $\propto |\varkappa|^{-1}$ exceeds the typical correlation length $\zeta_{corr}$. In dimensionless units, this sets a recombination time $t_{BS}\propto T_{max} \zeta_{corr}^{-1}$. Physically, only particles that remained close by during this time window are allowed to bind: let us consider two particles of rapidities $\lambda$ and $\lambda'$: On a timescale $t_{BS}$ they drift apart by a distance $t_{BS}(2\lambda-2\lambda')$ (we neglect the interaction on this time scale and approximate the effective velocity with the bare velocity $2\lambda$). Imposing the drift to be smaller than the bound state size at $t_{BS}$, i.e. $\zeta_{corr}$, we obtained a coarse-grained resolution in  rapidity space $\Delta \lambda \sim \zeta^2_{corr}/T_{max}$. Thus, the Dirac deltas in  rapidity space appearing in the maximum entropy ansatz will be broadened on a scale $\sim \Delta \lambda$, inducing overall $\propto T^{-1}_{max}$ corrections.
\begin{figure}[t!]
\begin{center}
\includegraphics[width=0.8\textwidth]{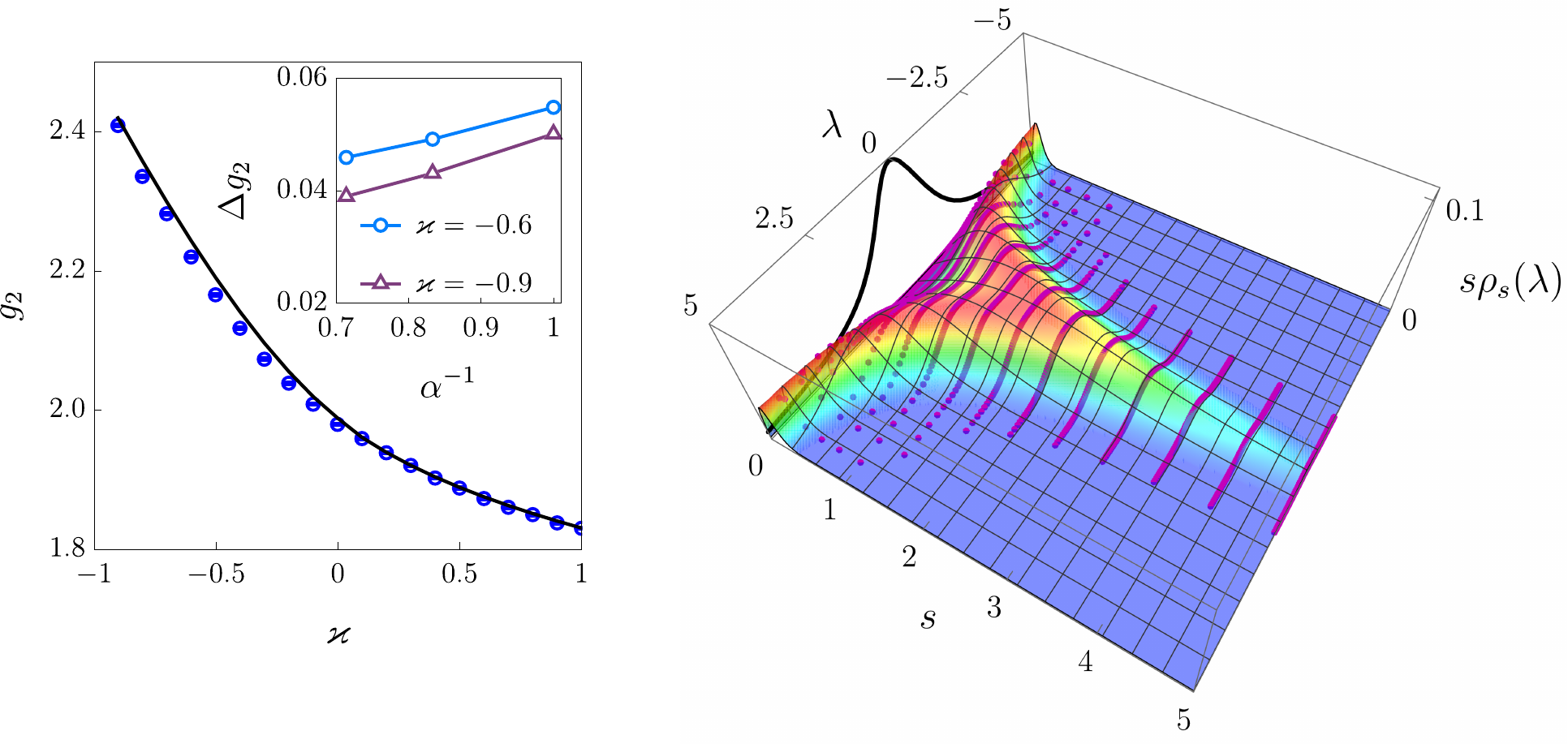}
\caption{\label{fig_th}
We initialize a thermal ensemble in the repulsive phase $\varkappa=1$ with inverse temperature $\beta=1$ and chemical potential $\mu=-0.1$, resulting in a density $\simeq 0.40$. The interaction is slowly ramped to $\varkappa=-1$ on a time $T_{max}=40$. 
The phase space is independently discretized within the $\lambda$ and $s$ space in an inhomogeneous fashion $\{s_j\}_{j=1}^{N_s}=S (20 (j/N_s)^2+(j/N_s))/21$ with $S=12$ and  $\{\lambda_j\}_{j=1}^{N_\lambda}=\Lambda (50 (j/N_\lambda)^{15}+(j/N_\lambda))/51$  with $\Lambda=150$. The non-linear discretization is tuned to compromise between the fat tails in rapidity space (which populates only small values of $s$) and the peak at $\lambda=0$ which extends up to large values of $s$. To further reduce the number of points, we tossed away those coordinates at which filling is basically zero during the whole evolution.
We consider three sets of data $(N_\lambda, N_s)=\{(\alpha_i 100, \alpha_i 15\}_{i=1}^3$ with $\alpha=\{1,6/5, 7/5\}$ and linearly extrapolate in $1/\alpha$.
Left: Monte Carlo data (markers) against extrapolated GHD data. Inset: GHD data at fixed interactions as a function of the extrapolation parameter $\alpha$.
Right: $s\rho_s$ is plotted at $\varkappa=0^-$, for comparison we show $\rho(\lambda)|_{\varkappa=0^+}$ (black line) and the points used for the best discretization (purple markers). 
}
\end{center}
\end{figure}
After having studied the approach to adiabaticity, we consider the protocol for different initial states focusing on $g_2$ (Fig. \ref{fig_gauss} $(b)$) and on the density-density correlation function \eqref{eq_corr_euler} (Fig. \ref{fig_gauss} $(c)$), where we find excellent agreement.
As expected, the gas is much more sensitive to attractive rather than repulsive interactions, as depicted in Fig. \ref{fig_gauss_rep}. There, we initialize the system at the non-interacting point and alternatively drive it within the attractive or repulsive phase. By retaining the same initial state, we explore the effects of the interactions. In particular, it is worthwhile to contrast the density correlation functions in the attractive phase with the repulsive phase (Fig. \ref{fig_gauss_rep} $(b)$ and $(c)$). By increasing $\varkappa$ to negative values, the particles feel stronger attraction and convert part of their kinetic energy into interactions. Hence, a strong signal appears at small rays (i.e. small velocities). On the contrary, repulsive interactions accelerate the particles to larger velocities and reduce the peak of the correlation function.
Finally, we address the original proposed protocol: in Fig. \ref{fig_th} we initialize the system in a thermal ensemble within the repulsive phase and adiabatically drive it to attractive interactions.
Ideally, thermal states are simply a different initial state which can be captured by our method, but in practice the discretization of the GHD equations appears more challenging. The reason is in the shape of $\rho(\lambda)|_{\varkappa=0^+}$, which features polynomially decaying tails at large rapitidities $\sim \lambda^{-2}$, while it is very peaked at small rapitidies. As a consequence, a very large cutoff in the rapidity space must be used, while in the attractive phase the cutoff along the $s$ direction greatly depends on $\lambda$.
These features, in addition to the need of a fine discretization in  rapidity space to correctly capture the integration kernels and the Euleran shift of the GHD equation, worsen the convergence. 
We experienced that a $\lambda-$dependent discretization of the $s$ space improves the convergence of the integral equations, but negatively affects the interpolation used in solving the GHD equation. Hence, we compromise with an independent, but inhomogeneous discretization of the $\lambda$ and the $s$ space: a judicious discretization choice is crucial to boost the convergence (see caption of Fig. \ref{fig_th}). Overall, a much denser discretization in the $\lambda$ space compared with the string space is needed for correctly describing the derivatives and integration kernels.
In Fig. \ref{fig_th} we obtained a good agreement by considering different discretizations and then extrapolating (see caption for more details), while in Figs. \ref{fig_gauss} and \ref{fig_gauss_rep} convergence was obtained without the need of any extrapolation and by employing a flat discretization in the $\lambda$ and $s$ space (roughly $\sim 800$ points have been used in the discretization).

\section{Conclusions and discussion}
\label{sec_conclusions}

In this work, we  constructed the integrable hydrodynamics of the one-dimensional attractive Non Linear Schroedinger equation.
We then applied these findings to adiabatic interaction changes from the repulsive to the attractive phase, observing soliton production and engineering a strongly excited nonequilibrium phase of matter. The integrability of the NLS stabilizes the attractive phase out of equilibrium, which would be highly unstable in the presence of thermalization.
We constructed the Generalized Hydrodynamics of the classical model by considering the proper semiclassical limit of the quantum Bose gas: the excellent agreement with ab-initio microscopic simulations confirms the correctness of the procedure, but this strategy leaves open some interesting questions.

The first one concerns the absence of radiative modes, as one could have naively expected from the inverse scattering analysis. 
Let us stress that in statistical ensembles in the thermodynamic limit the distinction between solitons and radiations is blurred since one can accommodate for an arbitrary number (divergent in the thermodynamic limit) of zeroes in the imaginary plane of the scattering data \eqref{eq_arep}.
In this perspective, one could interpret radiation as a condensation of light solitons (small binding energy) on the real axis of the spectral parameter. 
Within the $(\lambda,s)$ phase-space, this means that modes at $s\to 0$ should be macroscopically occupied, which is not the case within our protocol. In fact, when $\varkappa=0$ is crossed over in an adiabatic manner, solitons of arbitrary size $s$ are created with equal probability and $s=0$ does not stand out as a special point. It is natural to wonder if radiative modes can be excited by different protocols.

A second question concerns the classical interpretation of the Planck constant $h$. Indeed, in contrast with the repulsive phase, we showed that the Planck constant does not completely disappear after having taken the semiclassical limit and plays an important role as a regulator in the entropy of the soliton gas. Interestingly, the $h\to 0$ limit can be explicitly taken on the TBA equations obtained after entropy maximization \eqref{eq_max_en_nns}, but not for the entropy itself. Indeed, by replacing the non-singular parametrization \eqref{eq_barvarth} in the semiclassical Yang-Yang entropy \eqref{Eq_divergentEntropy}, at most $\sim \log s$ singularities (and thus are integrable) appear. As a consequence, the $\sim \log h$ term is not counterbalanced by a divergence in $\delta (h)$.
Since it does not seem possible to explicitly remove the Planck constant from the semiclassical Yang-Yang entropy, the same term should be recovered from a pure classical construction of the thermodynamics.
Retrospectively, we can give a phenomenological interpretation of the Planck constant as a infrared regularization of the classical theory. Indeed, NLS solitons have a width $\sim (s |\varkappa|)^{-1}$: when addressing the thermodynamics of a large, but finite system of length $L$, large solitons should be removed from the spectrum giving a cutoff $s|\varkappa|\gtrsim L^{-1}$. In our results, this role is taken by the cutoff $\delta(h)\propto h$ appearing in Eq. \eqref{Eq_divergentEntropy}. However, notice that the validity of this crude argument is harmed by the need of taking the non-interacting limit and furthermore it does not explain the appearance of the $\sim \log h$ in the entropy. 
A complete understanding of these points passes through a complete classical analysis of the attractive phase. We leave this interesting question open for the future.
Lastly, it would be interesting to extend our analysis to the sine-Gordon model (SG) \cite{doi:10.1142/9789812775344_0020,Mussardo:2010mgq,smirnov1992form,Faddeev:1987ph}. This relativistic field theory is integrable both at the quantum and classical level and its spectrum is understood in terms of topological excitations (kinks) and bound states of them (breathers). The interest is twofold: first, by changing the interaction one forms or destroys kink's bound states, in strict analogy with the NLS case. 
Secondly, the thermodynamics of the classical sine-Gordon, which in contrast to the NLS is stable also at equilibrium, has been thoroughly addressed in the literature also by means of semiclassical analysis \cite{PhysRevLett.62.708,PhysRevLett.56.2233,PhysRevLett.56.904}. On the other hand, the SG model and the NLS are closely connected, since the second can be obtained as the non-relativistic limit of the first \cite{PhysRevLett.103.210404,PhysRevA.81.043606,Calabrese_2014}. However, to the best of our knowledge, an analogue $\sim \log h$ regulator does not appear in the SG literature. It would be interesting to further investigate this point.

\section*{Acknowledgements}
We thank \v{Z}iga Krajnik, Enej Ilievski, Toma\v{z} Prosen, Benjamin Doyon, Oleksandr Gamayun and Andrea De Luca for fruitful discussions.
We are grateful to \v{Z}iga Krajnik, Enej Ilievski and Oleksandr Gamayun for useful comments on our manuscript.
RK and JSC acknowledge support from the European
Research Council (ERC) under ERC Advanced grant 743032 DYNAMINT.
AB acknowledges support from the Deutsche Forschungsgemeinschaft (DFG, German Research Foundation) under Germany's Excellence Strategy–EXC–2111–390814868.

\appendix
\section{The microscopic numerical simulations}
\label{app_micro_num}
The microscopic simulations are performed by discretizing the NLS $\phi(x)\to\phi_j$ on $N$ lattice sites and lattice space $a$, thus the system's length is $L=aN$. The limit of small lattice spacing is then numerically taken.
In this respect, one discretize the NLS Hamiltonian as
\be
H\to a\sum_j\Big\{ a^{-2}|\phi_{j+1}-\phi_j|^2+\varkappa |\phi_j|^4\Big\}\, ,
\ee
which results in the equation of motion
\be\label{ed_dis_eqm}
i\partial_t \phi_j=-\frac{1}{a^2}(\phi_{j+1}+\phi_{j-1}-2\phi_j)+2\varkappa |\phi_j|^2\phi_j\, .
\ee
Periodic boundary conditions are assumed. The evolution of the observables of interest is performed in two steps
\begin{enumerate}
    \item We randomly sample the initial condition $\phi_j(t=0)$ from a known probability distribution. We consider two cases which are sampled in different ways:
    \begin{enumerate}
        \item For initial values $\varkappa>0$ we consider thermal initial states of inverse temperature $\beta$ and chemical potential $\mu$. Therefore,  the target probability is $P[\{\phi_j\}]\propto \exp\left[-\beta (H-\mu a\sum_j |\phi_j|^2)\right]$. Boltzmann distributions of this form are efficiently sampled by means of Metropolis-Hasting algorithms: basically, one builds a stochastic random walk on the field configuration $\{\phi_j\}\to \{\phi_j'\}$, where the proposed update is accepted or rejected depending on $P[\{\phi'_j\}]/P[\{\phi_j\}]$. This process is ergodic in the phase space and, after a sufficient number of initial steps, the average over the probability $P$ can be replaced with an average over the Metropolis time. These techniques are standard and the reader can refer to Ref. \cite{doi:10.1063/1.1699114,10.1093/biomet/57.1.97,doi:10.1080/00031305.1995.10476177} for further details.
    \item By choosing $\varkappa=0$ as the initial point of the time evolution, we can conveniently initialize the state in arbitrary GGEs. Indeed, non-interacting GGEs are Gaussian states diagonal in the Fourier space
    \be
    \langle\tilde{\phi}_j\rangle=0\hspace{2pc}\langle \tilde{\phi}^*_j\tilde{\phi}_{j'}\rangle= \frac{ 2\pi}{L}\delta_{j,j'}\rho\left(\frac{2\pi}{L} j_{\text{mod}(N,-N/2)}\right)
    \ee
    where $\tilde{\phi}_j$ is the field in the discrete Fourier space
    \be
    \phi_j=\sum_{j'} e^{i2 \pi j j'/N} \phi_{j'}
    \ee
    and $j_{\text{mod}(N,-N/2)}$ is the periodic repetition of $j$ modulus $N$ computed within the interval $[-N/2,N/2]$. Above, $\rho(k)$ sets the initial root density in the limit $\varkappa\to 0^+$ and can be set at will.
    \end{enumerate}
    \item Each random initial field configuration $\phi_j(t=0)$ is then evolved according to the deterministic equation of motion \eqref{ed_dis_eqm}. In order to achieve stability, we trotterize the evolution on the time interval $\Delta t$ in two steps. In the first step, we diagonally evolve in the real space according to
    \be
    \phi_j(t+\Delta t/2)= e^{-i\Delta t \varkappa(t)|\phi_j(t)|^2}\phi_j(t)\, .
    \ee
    In the second step, we instead evolve in the Fourier space
    \be
    \tilde{\phi}_j(t+\Delta t)= e^{-i \frac{2\Delta t}{a^2}[1-\cos(2\pi j)]}\tilde{\phi}_j(t+\Delta t/2)\, .
    \ee
    This strategy explicitly enforces the particle number conservation. For a given lattice spacing $a$, we chose $\Delta t$ by checking energy conservation of an initial random configuration in the case of a constant interaction $\varkappa$. A posteriori, the convergence of the simulation is checked by considering different discretizations in space and time.
\end{enumerate}

\section{The numerical solution of the GHD equations}
\label{app_ghd_num}

We start with  discretizing the rapidity space and in case of attractive interactions  the string space, additionally. We here choose a possibly inhomogeneous, but independent discretization in the rapidity and string spaces $[\{\lambda_i\}_{i=1}^{N_{\lambda}}, \{s_\alpha\}_{\alpha=1}^{N_{s}}]$ which gives enough flexibility to compromise between numerical effectiveness and a sufficient resolution (further details on the concrete discretizaton can be found in the caption of Fig \ref{fig_th}). We moreover toss away points of the grid in the $(\lambda, s)$-plane where the filling fraction drops to zero.  
Within the attractive case, we use the non-singular reparametrization discussed in Eq. \eqref{eq_barvarth} and below.
\\
Next, several integral equations need to be discretized for which we take the dressing \eqref{eq_drbar} as an example in order to illustrate our proceedings. 
We use the grid in order to discretize Equation \eqref{eq_drbar} and get
\be
s_\alpha t_{s_\alpha}^{\bdr}(\lambda_i)=t_{s_\alpha}(\lambda_i)-\sum\limits_{\beta=1}^{N_s}\sum\limits_{j=1}^{N_{\lambda}}\Phi_{s_\alpha,s_{\beta}}(\lambda_i,\lambda_j)\bar{\vartheta}_{s_\beta}(\lambda_j) t_{s_\beta}^{\bdr}(\lambda_j)\, .
\ee
Here, $\Phi_{s_\alpha,s_\beta}(\lambda_i,\lambda_j)$ is the integration  kernel analytically integrated over  the domain surrounding $(\lambda_j,s_\beta)$ 
\be\label{eq_dis_intk}
\Phi_{s_\alpha,s_\beta}(\lambda_i,\lambda_j)=\int\limits_{\frac{1}{2}(s_\beta+s_{\beta-1})}^{\frac{1}{2}(s_\beta+s_{\beta+1})} \dd s' \int\limits_{\frac{1}{2}(\lambda_j+\lambda_{j-1})}^{\frac{1}{2}(\lambda_j+\lambda_{j+1})}\dd\lambda' s_\alpha^{-1}\varphi_{s_\alpha s'}(\lambda_i-\lambda')
\ee
The integral above can be analytically performed. Notice that the integrand becomes singular for small interactions, pointing out the necessity of the discretization Eq. \eqref{eq_dis_intk} when compared with a more naive midpoint rule. A similar strategy is used in computing the force term and in the repulsive phase as well.\\
To start the time evolution  determined by the GHD equations, we need to compute the initial state characterized by a root density. Within the repulsive phase, we initialize the system in a thermal state  which  we obtain upon solving the TBA equation \eqref{eq_effen_cl}  with a fixed chemical potential $\mu$. Alternatively, we can construct a root density in form of a Gaussian \eqref{eq_GGE_gauss} at the free point $\varkappa =0^+$ which comes in handy to control the fat tails in rapidity space that appear in the thermal state. 
To enter the attractive phase, we use the root density at $\varkappa=0^+$ as an input and maximize the entropy. We found that discretizing the integral equations \eqref{eq_max_en_nns} and then numerically solving them to yield exactly the input root density provides better accuracy rather than discretizing the analytical exact solution.
The integral appearing in Eq. \eqref{eq_max_en_nns} is computed by a similar discretization strategy as the one used for the $\varkappa\ne 0$ dressing.
In view of the constant cutoff in the string space employed in our discretization, we directly solve \eqref{eq_max_en_nns} by restricting the integral on a interval $[0,S]$ and then discretize it on a finite grid. This strategy, compared with discretizing directly \eqref{eq_max_en_cl_fill}, allows to obtain an exact particle number conservation also on a finite discretization grid.
The GHD equations are then solved in the filling space with the methods of characteristics.
First, one observes that the GHD equation for the filling fraction \eqref{eq_ghd_fill} can be exactly recast as an implicit translation
\be 
\vartheta_s(t + \Delta t, \lambda)=\vartheta_s\left(t, \lambda-\int\limits_{t}^{t+\Delta t}\dd \tau F_s^{\eff}(\lambda, \tau)\right)\, ,
\ee
where, for simplicity, we consider the homogeneous case. The time integral of the force is computed with a midpoint rule, obtaining a second order algorithm in the time displacement. For details see Refs. \cite{PhysRevLett.123.130602,mller2020introducing}.

\bibliography{biblio}

\end{document}